\RequirePackage{fix-cm}
\documentclass[smallextended]{svjour3}    
\smartqed 
\usepackage{lineno}
\usepackage{graphicx}
\usepackage{bm}
\usepackage{color}
\usepackage{booktabs}
\usepackage{amsmath}
\usepackage{multirow}
\usepackage{bmpsize}
\usepackage[utf8]{inputenc}
\usepackage{subfig}
\renewcommand\thefigure{\arabic{figure}} 
\renewcommand\thetable{\arabic{table}}

\newcommand*\patchAmsMathEnvironmentForLineno[1]{%
	\expandafter\let\csname old#1\expandafter\endcsname\csname #1\endcsname
	\expandafter\let\csname oldend#1\expandafter\endcsname\csname end#1\endcsname
	\renewenvironment{#1}%
	{\linenomath\csname old#1\endcsname}%
	{\csname oldend#1\endcsname\endlinenomath}}%
\newcommand*\patchBothAmsMathEnvironmentsForLineno[1]{%
	\patchAmsMathEnvironmentForLineno{#1}%
	\patchAmsMathEnvironmentForLineno{#1*}}%
\AtBeginDocument{%
	\patchBothAmsMathEnvironmentsForLineno{equation}%
	\patchBothAmsMathEnvironmentsForLineno{align}%
	\patchBothAmsMathEnvironmentsForLineno{flalign}%
	\patchBothAmsMathEnvironmentsForLineno{alignat}%
	\patchBothAmsMathEnvironmentsForLineno{gather}%
	\patchBothAmsMathEnvironmentsForLineno{multline}%
}

\begin{document}
	\title{Gradient-enhanced continuum models of healing in damaged soft biological tissues}
	
	\author{Yiqian He \and
		Di Zuo \and
		Klaus Hackl \and
		Haitian Yang \and
		S Jamaleddin Mousavi\and
		Stéphane Avril 
	}
	\institute{Yiqian He\and  Di Zuo \and Haitian Yang\at
		State Key Lab of Structural Analysis for Industrial Equipment, Department of Engineering Mechanics, Dalian University of Technology, Dalian 116024, P.R. China
		\and
		Klaus Hackl\at
		Mechanik – Materialtheorie, Ruhr-Universität Bochum, Bochum, Germany
		\and
		S Jamaleddin Mousavi \and Stéphane Avril*\at
		Ecole Nationale Supérieure des Mines de Saint-Etienne, CIS-EMSE, SAINBIOSE, F-42023 St. Étienne, France\\
		*Corresponding author E-mail: stephane.avril@mines-stetienne.fr
	}
	\maketitle
	\begin{abstract}
		Healing of soft biological tissue is the process of self-recovering or self-repairing the injured or damaged extracellular matrix (ECM). Healing is assumed to be stress-driven, with the objective of returning to a homeostatic stress metrics in the tissue after replacing the damaged ECM with new undamaged one. However, based on the existence of intrinsic length scale in soft tissues, it is thought that computational models of healing should be non-local. In the present study, we introduce for the first time two gradient-enhanced constitutive healing models for soft tissues including non-local variables. The first model combines a continuum damage model with a temporally homogenized growth model, where the growth direction is determined according to local principal stress directions. The second one is based on a gradient-enhanced healing model with continuously recoverable damage variable. Both models are implemented in the finite-element package Abaqus by means of a user subroutine UEL. Three two-dimensional situations simulating the healing process of soft tissues are modeled numerically with both models, and their application for simulation of balloon angioplasty is provided by illustrating the change of damage field and geometry in the media layer throughout the healing process.
	\end{abstract}
	
	\section{Introduction}
	\label{sec.1}
	Healing of soft biological tissue is the process of self-recovering or self-repairing the injured or damaged extracellular matrix (ECM). Healing is a complex biochemical and biomechanical process, usually divided into four stages: haemostasis, inflammation, proliferation and remodeling. These four stages were described in large details by Comellas et al. \cite{Comellas2016} and Cumming et al. \cite{Cumming2009}. It was reported that the first three stages (from haemostasis to proliferation) may last several weeks, and the final stage of remodeling may last from weeks to years. This last stage consists in a continuous turnover (synthesis and degradation) of the ECM simultaneously with the production of scar tissue. 
	
	Computational modeling can provide insight into healing of soft tissues both at short term and long term. Numerical simulation of healing in soft tissues has been a topic of intense research. Tepole and Kuhl \cite{BuganzaTepole2013} and Valero et al. \cite{Valero2015} provided a comprehensive review of computational models of dermal wound healing. Generally, there are two types of approaches. The first type focuses on the underlying cellular and biochemical mechanisms based on continuum or hybrid discrete/continuum approaches, including the simulation of wound contraction \cite{Javierre2009,Valero2015a} and angiogenesis \cite{Schugart2008}. Another type of approaches, more phenomenological, focuses on the change of material properties in the tissue during the remodelling phase. 
	
	Important mechanisms involved in soft tissue healing, such as collagen fiber reorientation and collagen turnover, were modeled using growth and remodeling (G$\&$R). A large number of G$\&$R computational approaches exist, among which the constrained mixture theory which was introduced by Humphrey and Rajagopal \cite{HUMPHREY2002} about 20 years ago and has been employed by many others \cite{Valentin2012,Valentin2013,Famaey2018}. The computational cost of this approach was significantly reduced by a temporally homogenized technique proposed by Cyron et al. \cite{Cyron2016}. 
	
	Recently, Comellas et al. \cite{Comellas2016} developed a homeostatic-driven turnover remodeling model for healing in soft tissues based on continuum damage mechanics (CDM). In this approach, the healing process was simulated by a continuously recoverable damage variable \cite{Comellas2016}. 
	
	Intrinsic length scales, such as the length of collagen fibers, are physically inherent to soft tissues and numerical models of healing should consider them through non-local approaches. Moreover, mesh dependency is a traditional issue in damage models that non-local approaches are able to overcome \cite{Dimitrijevic2008,Dimitrijevic2011,Waffenschmidt2014}. However, there are no non-local computational models of healing in the literature to the authors’ best knowledge, and the effects of intrinsic length scales in healing are still unknown.
	
	From the viewpoint of continuum damage mechanics, local continuum damage models have a major drawback: their solutions are significantly mesh-dependent, with vanishing of the localized damage zone when the mesh is refined \cite{Waffenschmidt2014,Mousavi2018}. 
	
	First non-local damage models of soft tissues were introduced by Waffenschmidt et al. \cite{Waffenschmidt2014}, using a gradient-enhanced large-deformation continuum damage model based on the previous work of Dimitrijevic and Hackl \cite{Dimitrijevic2008,Dimitrijevic2011}. In this approach, the local free energy function is enhanced by a gradient-term containing the gradient of an additional non-local damage variable, and a penalization term is also introduced to ensure equivalence between the local and non-local damage variables. Ferreira et al. \cite{Ferreira2017} also presented an integral-type non-local averaging damage model for anisotropic hyperelastic materials. Despite this state of the art in non-local damage modeling, healing of soft tissues remains a frontier in non-local continuum mechanics.
	
	In this paper, we introduce for the first time two gradient-enhanced constitutive healing models for soft tissues including non-local variables in a similar fashion as in previous work from Dimitrijevic and Hackl \cite{Dimitrijevic2008,Dimitrijevic2011} and Waffenschmidt et al. \cite{Waffenschmidt2014}. By virtue of the proposed model, an intrinsic length scale parameter is for the first time included in a healing model with a gradient-enhanced term, and a non-local variable is introduced with a penalization term to reduce mesh dependency.
	
	The first non-local healing model combines the non-local continuum damage model with a temporally homogenized G$\&$R model. Damage is modeled with the gradient-enhanced approach, and a term of mass production is introduced to model mass variations due to tissue production. There exist a variety of growth models, from surface to volume growth, taking into account mass variations in biological materials as described in \cite{Ganghoffer2005,Ganghoffer2010,Ganghoffer2012,Ganghoffer2017,Ganghoffer2010a}. In this work, a temporally homogenized growth model is used based on the work from Cyron et al. \cite{Cyron2016}, permitting significant reduction of the computational cost compared to original work from \cite{HUMPHREY2002}. In this temporally homogenized growth model, the rate of mass production satisfies a homeostasis-driven governing equation. Mass production induces inelastic deformations which are modeled in a similar fashion as in plasticity \cite{Rodriguez1994}. We assume that the growth direction is aligned with the direction of the first principal stress. 
	
	The second non-local healing model is based on the healing model proposed by Comellas et al. \cite{Comellas2016} that we turned into a gradient-enhanced version. In this model, the healing is simulated by turning the damage variable into a recoverable variable. The process of damage recovery is controlled by the healing rate, and can be integrated numerically by a finite difference scheme. 
	
	Both models are implemented in the finite-element package Abaqus by means of a user subroutine UEL. In the following, the general gradient-enhanced G$\&$R healing model is developed in Section 2. Section 3 provides two specific gradient-enhanced healing models, including the details of equations for the rate and direction of growth and the evolution of damage. Section 4 outlines the process of numerical implementation of the proposed methods. Three examples are illustrated in Section 5 with the aim of verifying these models. Finally, conclusions are given in Section 6.
	
	\section{General equations for gradient-enhanced healing models}
	\label{sec.2}
	\subsection{Basic kinematics}
	\label{sec.2.1}
	Let ${{x}} = {\bm{\varphi }}({\bf{X,}}t)$ describe deformations of a body from referential positions ${\bf{X}} \in \kappa (0)$ to their actual counterparts ${{x}} \in \kappa (t)$. Within this framework, the deformation gradient is defined as
	\begin{equation}
		\label{equ.1}
		{\bf{F}} = {\nabla _{\bf{X}}}{\bm{\varphi }}
	\end{equation}
	
	Accordingly, reference volumes ${\rm{d}}V$ and current volumes ${\rm{d}}v$ are related such as 
	\begin{equation}
		\label{equ.2}
		{\rm{d}}v = \det ({\bf{F}}){\rm{d}}V{\rm{= }}J{\rm{d}}V
	\end{equation}
	where $J$ is the Jacobian of the deformation (determinant of {$\bf{F}$}).
	
	Growth is a process of mass production or removal, whereby volumes may change inelastically. This is captured by an inelastic deformation gradient ${{\bf{F}}_g}$. Therefore the total deformation at any time $t$ is
	\begin{equation}
		\label{equ.3}
		{\bf{F}}(t){\rm{= }}{{\bf{F}}_e}(t){{\bf{F}}_g}(t)
	\end{equation}
	
	\subsection{Gradient-enhanced healing model}
	\label{sec.2.2}
	
		The general strain energy function per unit reference volume at each G$\&$R time is assumed as
		\begin{equation}
			\label{equ.4}
			\psi (t){\rm{= }}H(t)\hat \psi ({{\bf{F}}_e}(t))
		\end{equation}
		where $\hat \psi ({{\bf{F}}_e}(t))$ is the original (undamaged) strain energy depending on the elastic deformation ${\bf F}_e$, $H(t)$ is to a time-dependent function to describe the level of healing and has different forms for different healing models in the following section.
	
	Following the approach of Dimitrijevic and Hackl \cite{Dimitrijevic2008,Dimitrijevic2011}, a gradient-enhanced non-local free energy function is added to the energy given in Equation (4),
	\begin{equation}
		\label{equ.5}
		\psi (t){\rm{= }}H(t)   \hat \psi ({{\bf{F}}_e}(t)) + \frac{{{c_d}}}{2}{\left\| {{\nabla _{\bf{X}}}\phi } \right\|^2} + \frac{{{\beta _d}}}{2}{\left[ {\phi  - {\gamma _d}d} \right]^2}
	\end{equation}
	
	In Equation (5), $c_d$ represents the gradient parameter that defines the degree of gradient regularization and the internal length scale. Comparing Equations (4) and (5), two additional terms are added, introducing the 3 following variables:
		
		- the variable field $\phi$ , which transfers the values of the damage parameter across the element boundaries to make it non-local in nature, 
		
		- the energy-related penalty parameter $\beta_d$ which approximately enforces the local damage field and the non-local field to coincide, 
		
		- parameter $\gamma_d$ which is used as a switch between the local and enhanced model.
	\section{Two gradient-enhanced healing models}
	\label{sec.3}
	\subsection{Gradient-enhanced healing model based on G$\&$R}
	\label{sec.3.1}
	In this section, a new gradient-enhanced healing model based on G$\&$R is presented inspired by Valentín et al. \cite{Valentin2013}. The strain energy function per unit reference volume at each G$\&$R time $t$ is assumed as
		\begin{equation}
			\label{equ.6}
			\psi_1 (t){\rm{= }}H_1(t)   \hat \psi ({{\bf{F}}_e}(t)) + \frac{{{c_d}}}{2}{\left\| {{\nabla _{\bf{X}}}\phi } \right\|^2} + \frac{{{\beta _d}}}{2}{\left[ {\phi  - {\gamma _d}d} \right]^2}
		\end{equation}
		where
		\begin{equation}
			\label{equ.7}
			H_1(t){\rm{= }}f(d)\frac{{{\rho _0}}}{{\rho (t)}}Q(t) + \frac{{{\rho _g}(t)}}{{\rho (t)}}
	\end{equation}
	
	In Equations (7) $\rho_0$ is mass density per unit reference volume at $t=0$, just prior to the beginning of G$\&$R, $\rho_g(t)$ denotes the change of mass density computed by ${\rho _g}(t) = \rho (t) - {\rho _0}$, and $\rho_g(t)$ are caused by G$\&$R only and induce inelastic deformations whereas the motion induces elastic motions. $Q(t) \in \left[ {0,1} \right]$ is the mass fraction that was present at $t=0$ that survives to time $t$ \cite{Valentin2013} and $f(d)$ represents a function of damage variable $d$ that measures the material stiffness loss and satisfies the conditions material stiffness loss and satisfies the conditions
	\begin{equation}
		\label{equ.8}
		f(d) \quad :\quad  {\Re ^ + } \to (0,1]\left\{{f(0) = 1, \mathop {\lim f(d) = 0}\limits_{d \to \infty } } \right\} \quad  {{\rm with} \quad f(d) \in \left[ {0,1} \right]}
	\end{equation}
	
	It is noted that density variations $\rho_g(t)$  are caused by G$\&$R and induce inelastic deformations, whereas the elastic deformation gradient  ${\bf{F}}_e(t)$ satisfies Equation (3) and Equation (4). 
		
		According to Braeu et al. \cite{Braeu2017}, the deformation caused by growth is regarded as an inelastic deformation, where the change of volume is related to a change in mass. Hence, the rate of inelastic deformation gradient ${\bf{\dot F}}_g$  is obtained as in Braeu et al. \cite{Braeu2017}
	\begin{equation}
		\label{equ.9}
		{{\bf{\dot F}}_g}{\rm{= }}\frac{{\dot \rho (t)}}{{\rho (0)\left| {{{\bf{F}}_g}} \right|\left[ {{{({{\bf{F}}_g})}^{- {\rm{T}}}}:{\bf{B}}} \right]}}{\bf{B}}
	\end{equation}
	where the second-order tensor $\bf B$  defines the growth direction and is normalized without loss of generality such that $tr(\bf B)=1$. 
	
	The Davis’ law \cite{davis1867conservative} suggests that perturbations from a preferred homeostatic state in soft collagenous tissues are answered by biological G$\&$R processes aimed to restore normalcy. The Davis’s law can be invoked to justify anisotropic growth, as adding mass in directions normal to the maximum principal stress will automatically reduce the stress value and make it converge back to the homeostatic value \cite{Cyron2017,Menzel2005}. A very good case illustrating this effect is related to the thickening of arteries due to hypertension. Indeed, many observations showed that arteries tend to thicken in response to sustained increases in blood pressure (i.e., hypertension) \cite{Saez2014}. Hence, we assume the growth direction is aligned with the direction of the first principal stress. For instance, in two-dimensional cases, if   is the orientation of the first principal stress, the growth direction tensor $\bf B$ in Equation (9) can be expressed as
	\begin{equation}
		\label{equ.10}
		{\bf{B}}{\rm{= }}\left[ {\begin{array}{*{20}{c}}
				{{{\cos }^2}{\theta _p}}&0\\
				0&{{{\sin }^2}{\theta _p}}
		\end{array}} \right]
	\end{equation}
	
	To determine the rate of mass production caused by growth ${\dot \rho _g}(t)$  in Equation (\ref{equ.7}), two models are considered in this paper:
	
	\subsubsection{G$\&$R constant model}
	\label{sec.3.3.1}
	In the \emph{G}\&\emph{R constant model}, the mass production is assumed to be constant during the healing process as 
	\begin{equation}
		\label{equ.11}
		{k_g} \cdot {\rho _0} \cdot Q(t) + {\rho _g}(t) = {\emph{{const}}}
	\end{equation}
	where $k_g$  is the healing fraction to denote the percentage of mass before the healing to participate to mass balance. 
	
	Therefore, the $\dot\rho_g(t)$  is obtained by computing the time derivative such as
	\begin{equation}
		\label{equ_12}
		{\dot \rho _g}(t) =  - {k_g} \cdot {\rho _0} \cdot \dot Q(t)
	\end{equation}
	
	Considering total mass density $\rho (t) = {\rho _0} + {\rho _g}(t)$, so the rate of total mass density is
	\begin{equation}
		\label{equ.13}
		\dot \rho (t) = (1 - {k_g}){\rho _0} \cdot \dot Q(t)
	\end{equation}
	
	The total mass density $\rho(t)$  at time step $n+1$  can be obtained by the finite difference scheme 
	\begin{equation}
		\label{equ.14}
		\rho ({t_{{\rm{n + 1}}}}){\rm{= }}(1 - {k_g}){\rho _0} \cdot \dot Q({t_{\rm{n}}}) \cdot \Delta t{\rm{+ }}\rho ({t_{\rm{n}}})
	\end{equation}
	where $\Delta t$ is the time step.
	
	According to Equations (\ref{equ.11}) and (\ref{equ.14}), the mass densities  $\rho_g(t)$ and $\rho(t)$ in Equation (7) are determined. 
	
	\subsubsection{G$\&$R homeostatic model}
	\label{sec.3.1.2}
	
	In the \emph{G}\&\emph{R homeostatic model}, the rate of mass production is mediated by the current stress as proposed by Braeu et al. \cite{Braeu2017}
	\begin{equation}
		\label{equ.15}
		{\dot \rho _g}(t){\rm{= }}{\rho _g}(t){{\bf{K}}_\sigma }:\left( {{{\bm{\sigma }}_R}} - {{\bm{\sigma }}_h} \right) + \dot D(t)
	\end{equation}
	where $\bf K_\sigma$   is a gain-type second-order tensor, for two-dimensional case, it is assumed that
	\begin{equation}
		\label{equ.16}
		{{\bf{K}}_\sigma }{\rm{= }}\left[ {\begin{array}{*{20}{c}}
				{{k_\sigma }}&0\\
				0&{{k_\sigma }}
		\end{array}} \right]
	\end{equation}
	and $\dot{D}(t)$   is a generic rate function for additional deposition that is not stress mediated(describing additional deposition or damage processes affecting the net mass production driven by other factors such as chemical degradation and/or mechanical fatigue processes), ${{\bm{\sigma }}_R} = {{\bf{R}}^{\rm{T}}}{\bm{\sigma R}}$ is the co-rotated Cauchy stress tensor with the orthonormal rotation tensor  $\bf R$ in polar decomposition and $\bm{\sigma}_h$  denote the homeostatic stress.
	
	The mass production  $\rho_g(t)$ at time step $n+1$ can be obtained by the finite difference scheme in the absent of   $\dot{D}(t)$ for simplicity 
	\begin{equation}
		\label{equ.17}
		{\rho _g}({t_{{\rm{n + 1}}}}){\rm{= }} {{\rho _g}({t_{\rm{n}}}){{\bf{K}}_\sigma }:\left( {{{\bm{\sigma }}_R} - {{\bm{\sigma }}_h}} \right)}\cdot \Delta t{\rm{+ }}{\rho _g}({t_{\rm{n}}})
	\end{equation}
	
	Accordingly, total mass density $\rho(t)$ at time step $n+1$ can be obtained by the finite difference scheme 
	\begin{equation}
		\label{equ.18}
		\rho ({t_{{\rm{n + 1}}}}){\rm{= }} {{\rho _g}({t_{\rm{n}}}){{\bf{K}}_\sigma }:\left( {{{\bm{\sigma }}_R} - {{\bm{\sigma }}_h}} \right)} \cdot \Delta t{\rm{+ }}{\rho _g}({t_{\rm{n}}}) + {\rho _0} \cdot Q({t_{{\rm{n + 1}}}})
	\end{equation}
	
	\subsection{Non-local Comellas mode}
	\label{sec.3.2}
	In this section, another gradient-enhanced healing model is established based on the healing model proposed by Comellas et al. \cite{Comellas2016}, in which the effective damage $D_{eff}$ is assumed as a recoverable variable in the process of healing. In this paper, we apply similar constitutive equations into a gradient-enhanced framework. Here only some key equations for healing process are written, the readers can refer to literature \cite{Comellas2016} for detailed equations.
	
	The strain energy function per unit reference volume is written such as
		\begin{equation}
			\label{equ.19}
			\psi_2 (t){\rm{= }}H_2(t)   \hat \psi ({{\bf{F}}_e}(t)) + \frac{{{c_d}}}{2}{\left\| {{\nabla _{\bf{X}}}\phi } \right\|^2} + \frac{{{\beta _d}}}{2}{\left[ {\phi  - {\gamma _d}d} \right]^2}
		\end{equation}
		where 
		\begin{equation}
			\label{equ.20}
			{H_2}(t){\rm{= 1}} - {D_{eff}}\left( t \right)
		\end{equation}  
		
		In Equation (19), the second term is to introduce the gradient parameter $c_d$ that defines the degree of gradient regularization and the internal length scale. In order to make the model non-local, the third term is used for penalizing the difference between the damage field $d$ and the nonlocal variable field  $\phi$.
	
	According to Comellas et al. \cite{Comellas2016}, the effective damage $D_{eff}$  is assumed to a recoverable variable, and it's rate 
	\begin{equation}
		\label{equ_21}
		{\dot D_{eff}}{\rm{= }}\dot D - \dot R 
	\end{equation}
	where $\dot{D}$ is the rate of explicit Kachanov-like mechanical damage variable $D=f(d)$, and $\dot{R}$   is the healing rate given as
	\begin{equation}
		\label{equ.22}
		\dot R = \dot \eta \left\langle {{D_{eff}} - \xi } \right\rangle 
	\end{equation}
	where $\left\langle  \cdot  \right\rangle$   represents the Macaulay brackets,  $\dot{\eta}$ is a function that regulates how fast healing occurs and $\xi$   defines the percentage of stiffness that is not recovered at the end of the healing process. 
	
	The effective damage at time step $n+1$ can be obtained by the finite difference scheme proposed by Comellas et al. \cite{Comellas2016} as
	\begin{equation}
		\label{equ.23}
		D_{eff}^{n + 1} = (D_{eff}^n + \Delta D + \dot \eta \xi \Delta t)/(1 + \dot \eta \Delta t)
	\end{equation}
	
	\subsection{Total potential energy and variational formulation}
	\label{3.3}
	The potential energy can be written as \cite{Dimitrijevic2008,Dimitrijevic2011}
	\begin{equation}
		\label{equ.24}
		\Pi  = \int\limits_\Omega  \psi  {\rm{d}}V - \int\limits_\Omega  {{\bf{\bar B}} \cdot {\bf{\varphi }}} {\rm{d}}V - \int\limits_{\partial \Omega } {{\bf{\bar T}} \cdot {\bf{\varphi }}} {\rm{d}}V
	\end{equation}
	where  $\bar{\bf B}$ denotes the body force vector per unit reference volume and $\bar{\bf T}$   characterizes the traction vector per unit reference surface area. $\Omega$ represents the reference volume, and  $\partial \Omega$ is the surface boundary of $\Omega$.
	
	Minimization of the potential energy with respect to the primal variables $\bm\varphi$ and $\phi$ results in a system of equations that have to be zeroed globally
	\label{equ_20}
	\begin{equation}
		\int\limits_\Omega  {{\bf{P}}:{\nabla _{\bf{X}}}\delta {\bm{\varphi }}} {\rm{d}}V - \int\limits_\Omega  {{\bf{\bar B}} \cdot \delta {\bm{\varphi }}} {\rm{d}}V - \int\limits_{\partial \Omega } {{\bf{\bar T}} \cdot \delta {\bm{\varphi }}} {\rm{d}}V = 0
	\end{equation}
	\begin{equation}
		\label{equ_21}
		\int\limits_\Omega  {{\bf{Y}}:{\nabla _{\bf{X}}}\delta \phi } {\rm{d}}V - \int\limits_\Omega  {Y\delta \phi } {\rm{d}}V = 0
	\end{equation}
	where $\bar{\bf B}$ is the body force vector, and $\bf P$ is the first Piola-Kirchhoff stress tensor.
	
	The vectorial damage quantity $\bf Y$ and the scalar damage quantity $\emph Y$ are defined such as
	\begin{equation}
		\label{equ_22}
		{\bf{P}} = {\partial _{{\bf F}_e}}\psi ,{\bf{Y}} = {\partial _{{\nabla _{\bf X}}\phi }}\psi ,Y = {\partial _\phi }\psi 
	\end{equation}
	
	Accordingly, the spatial quantities are given by
	\begin{equation}
		\label{equ_23}
		{\bm{\sigma }} = {\bf{P}} \cdot {\rm{cof(}}{{\bf{F}}^{- 1}}{\rm{)}},y = {\bf{Y}} \cdot {\rm{cof(}}{{\bf{F}}^{- 1}}{\rm{)}},y = {J^{- 1}}Y
	\end{equation}
	where the factor defined as ${\rm{cof(}}{\bf{F}}{\rm{) = }}J{{\bf{F}}^{- T}}$.
	
	\subsection{Damage evolution}
	\label{sec.3.4}
	The evolution of the damage variable $d$ can be found in the works by Dimitrijevic and Hackl \cite{Dimitrijevic2008,Dimitrijevic2011} and Waffenschmidt et al. \cite{Waffenschmidt2014}, here only some key equations are outlined.
	
	Following standard thermodynamic consideration of Dimitrijevic and Hackl \cite{Dimitrijevic2008,Dimitrijevic2011}, damage conjugate $q$ is defined as
	\begin{equation}
		\label{equ.29}
		q =  - \frac{{\partial \psi }}{{\partial d}}
	\end{equation}
	
	The damage condition at any time of the loading process is based on an energy-release rate threshold condition and corresponds to the model of Simo and Ju \cite{Simo1987}
	\begin{equation}
		\label{equ.30}
		{\Phi _d}{\rm{= }}q - {r_1} \le 0
	\end{equation}
	
	Based on the postulate of maximum dissipation, the differential equation of the evolution of damage variable is subjected to Kuhn-Tucker optimality conditions \cite{Dimitrijevic2008,Dimitrijevic2011}
	\begin{equation}
		\label{equ.31}
		\dot d = \dot \kappa \frac{{\partial {\Phi _d}}}{{\partial q}},\quad \dot \kappa  \ge 0,\quad {\Phi _d} \le 0, \quad \dot \kappa {\Phi _d}{\rm{= }}0
	\end{equation}
	
	\section{Finite element discretisation}
	\label{sec.4}
	In order to approach the process of replacing the damaged soft tissue with new undamaged, FE computation is also divided into two stages, i.e. the damage process and the healing process. This section only derives the implementation of FE for the healing process. For the detailed process of FE computation process, the readers can refer the work by Waffenschidt et al. \cite{Waffenschmidt2014}. 
	
	Following the works of Dimitrijevic and Hackl \cite{Dimitrijevic2008,Dimitrijevic2011} and Waffenschidt et al. \cite{Waffenschmidt2014}, a quadratic serendipity interpolation is used for both the geometry $\bf X$ and the field variables $\bm \varphi$, and a bilinear interpolation is used for the non-local field $\phi$. According to the isoparametric concept, these interpolations are written as
	\begin{equation}
		\label{equ_32}
		{{\bf{X}}^h} = \sum\limits_{I = 1}^{n_{en}^{\bm{\varphi}} } {{N_I}\left( \xi  \right)} {{\bf{X}}_I},\quad {{\bm{\varphi }}^h} = \sum\limits_{I = 1}^{n_{en}^{\bm{\varphi}}} {{N_I}\left( \xi  \right)} {{\bm{\varphi }}_I},\quad {\phi ^h} = \sum\limits_{I = 1}^{n_{en}^\phi } {{N_I}\left( \xi  \right)} {\phi _I}
	\end{equation}
	where $\xi$ denotes the coordinates in the reference element, ${n_{en}^{\bm{\varphi}}}$ and ${n_{en}^{\phi}}$ are the displacement-nodes and non-local-damage-nodes per element, respectively. 
	
	For the healing process, at a loading time $t$, an incremental scheme based on Newton's method is applied \cite{Dimitrijevic2008,Dimitrijevic2011}
	\begin{equation}
		\label{equ.33}
		{\left[ {\begin{array}{*{20}{c}}
					{{{\bf{R}}_{\bm{\varphi }}}}\\
					{{{\bf{R}}_\phi }}
			\end{array}} \right]^{i}}{+ }{\left[ {\begin{array}{*{20}{c}}
					{{{\bf{K}}_{{\bm{\varphi \varphi }}}}}\quad & {{{\bf{K}}_{{\bm{\varphi }}\phi }}}\\
					{{{\bf{K}}_{\phi {\bm{\varphi }}}}}\quad & {{{\bf{K}}_{\phi \phi }}}
			\end{array}} \right]^{i}} \cdot {\left[ {\begin{array}{*{20}{c}}
					{\Delta {\bm{\varphi }}}\\
					{\Delta \phi }
			\end{array}} \right]^{{i + 1}}}{= }\left[ {\begin{array}{*{20}{c}}
				{\bf{0}}\\
				{\bf{0}}
		\end{array}} \right]
	\end{equation}
	where 
	\begin{equation}
		\label{equ.34}
		{{\bf{K}}_{{\bm{\varphi \varphi }}}}{\rm{= }}\int_\Omega  {\nabla _{{x}}^TN \cdot \left[ {{{\bf{C}}_h}(t)} \right]{\kern 1pt} {\kern 1pt} {\kern 1pt}  \cdot } {\kern 1pt} {\kern 1pt} {\nabla _x}N{\kern 1pt} {\kern 1pt} {\rm{d}}v + \int_\Omega  {\left[ {\nabla _x^TN \cdot {\bm{\sigma }}{\kern 1pt} {\kern 1pt} {\kern 1pt}  \cdot {\nabla _x}N} \right]{\kern 1pt} {\kern 1pt} {\kern 1pt} {\bf{I}}{\kern 1pt} {\kern 1pt} } {\kern 1pt} {\rm{d}}v
	\end{equation}
	\begin{equation}
		\label{equ.35}
		{{\bf{K}}_{{\bf{\varphi }}\phi }}{\rm{= }}\int_\Omega  {\nabla _x^TN \cdot \frac{{d{\bm{\sigma }}}}{{d\phi }} \cdot N{\rm{d}}v} {\kern 1pt} {\kern 1pt} {\kern 1pt} 
	\end{equation}
	\begin{equation}
		\label{equ.36}
		{{\bf{K}}_{\phi {\bf{\varphi}}}}{\rm{= }}\int_\Omega  {N_{}^T \cdot 2\frac{{dy}}{{d{\bf{g}}}} \cdot \nabla _x^TN{\rm{d}}v} {\kern 1pt} {\kern 1pt} {\kern 1pt} 
	\end{equation}
	\begin{equation}
		\label{equ.37}
		{{\bf{K}}_{\phi \phi }}{\rm{= }}\int_\Omega  {N_{}^T \cdot \frac{{dy}}{{d\phi }} \cdot N{\kern 1pt} {\kern 1pt} {\rm{d}}v} {\kern 1pt} {\kern 1pt} {\rm{+ }}{\kern 1pt} {\kern 1pt} {\kern 1pt} \int_\Omega  {\nabla _x^TN \cdot {\kern 1pt} \frac{{d{\bf{y}}}}{{d\phi }} \cdot {\kern 1pt} {\kern 1pt} \nabla _x^TN{\rm{d}}v} 
	\end{equation}
	
	In above equations the tangent terms ${d{\bm{{\sigma}}}/{d{\phi}}}$, $2{dy}/{d{\bf g}}$, ${dy}/{d{\phi}}$ and ${d{\bf{y}}/{d{\phi}}}$ are the same with the damage process as in the work by Waffenschidt et al. \cite{Waffenschmidt2014}, ${{\bf{C}}_h}(t)$ is a new time-dependent tangent stress-strain matrix in the damage and healing process given as
	\begin{equation}
		\label{equ.38}
		{{\bf{C}}_h}(t){\rm{= }}H{\rm{(}}t{\rm{)}} \cdot {{\bf{C}}_e}
	\end{equation}
	where the ${\bf C}_e$ is the elasticity tensors for undamaged material, $H(t)$ defined in Equation (7) is to describe the level of healing and has different forms for the $non$-$local \ G\&R \ healing \ model$ and the $non$-$local \ Comellas \ model$ as introduced in the following section.
	
	\subsection{G$\&$R healing model with gradient-enhanced damage}
	\label{4.1}
	For the $non$-$local \ G\&R \ healing \ model$, the form of $H(t)$ is determined by the choice of model for mass production.
	
	If the \emph{G}$\&$\emph{R constant models} with finite difference scheme in the time domain is used, $H(t)$ at the $(n+1)th$ time step by substituting Equations (12) and (14) into Equation (7)
	\begin{equation}
		\label{equ.39}
		H{\rm{(}}{t_{n + 1}}{\rm{) = }}\frac{{{\rho _0}}}{{\rho ({t_{n + 1}})}}f(d)Q({t_{n + 1}}) + \frac{{{\kern 1pt} (1 - {k_g}){\rho _0} \cdot \dot Q({t_{\rm{n}}}) \cdot \Delta t{\rm{+ }}{\rho _g}({t_{\rm{n}}})}}{{\rho ({t_{n + 1}})}}{\kern 1pt} {\kern 1pt} 
	\end{equation}
	
	If the \emph{G}\&\emph{R homeostatic model} with finite difference scheme in time domain is  obtained by substituting Equations (15) and (18) into Equation (7)
	\begin{equation}
		\label{equ.40}
		\begin{aligned}
			H(t_{n + 1})=  &\frac{{{\rho _0}}}{{\rho ({t_{n + 1}})}}f(d)Q({t_{n + 1}})\\
			&+ \frac{{{\kern 1pt} {\kern 1pt}  {{\rho _g}({t_{\rm{n}}}){{\bm{K}}_\sigma }:\left( {{{\bm{\sigma }}_R} - {{\bm{\sigma }}_h}} \right)} \cdot \Delta t{\rm{+ }}{\rho _g}({t_{\rm{n}}}) + {\rho _0} \cdot Q({t_{{\rm{n + 1}}}})}}{{\rho ({t_{n + 1}})}}{\kern 1pt}
		\end{aligned}
	\end{equation}
	
	The elastic deformation obtained from the Equation (32) and the inelastic deformation due to G\&R from the Equation (9), finally the total deformation is gived by
	\begin{equation}
		\label{equ.41}
		{\bf{F}}({t_{n + 1}}){\rm{= }}{\nabla _{\bf{X}}}{\bm{\varphi }}({t_{n + 1}}) \cdot {{\bf{F}}_g}({t_{n + 1}})
	\end{equation}
	
	 In Equation (41), the first factor on the right hand side ${\nabla _{\bf{X}}}{\bf{\varphi }}({t_{n + 1}})$ refers to the elastic deformation, and the left-hand side ${\bf{F}}({t_{n + 1}})$ is the total deformation gradient.
	
	\subsection{Non-local Comellas model}
	\label{sec.4.2}
	
	For the \emph{non-local Comellas model}, 
	\begin{equation}
		\label{equ.42}
		H{\rm{(}}{t_{n + 1}}{\rm{) = 1}} - ({D_{eff}}\left( {{t_n}{\rm}} \right) + \Delta D + \dot \eta \xi \Delta t)/(1 + \dot \eta \Delta t)
	\end{equation}
	
	Since the deformation due to growth is not considered in \emph{non-local Comellas model}, the total deformation is given without ${{\bf{F}}_g}({t_{n + 1}})$  by
	\begin{equation}
		\label{equ.43}
		{\bf{F}}({t_{n + 1}}){\rm{= }}{\nabla _{\bf{X}}}{\bm{\varphi }}({t_{n + 1}})
	\end{equation}
	
	{The process of numerical implementation is provided in Table 1.} Examples for these expressions are given in the following sections for specific applications.
	
	\section{Numerical examples }
	\label{sec.5}
	The gradient-enhanced model and the different healing models are incorporated within the commercial finite-element software Abaqus/Standard by means of a user element subroutine (UEL) and the 2D examples are solved as plane strain problems. They were applied in 3 different situations described in the following subsections. In all of them a simple damage function $f(d)=e^{-d}$ is used. 
	
	\subsection{Uniaxial tension}
	\label{sec.5.1}
	A square plate with 1cm edge length is subjected to a displacement-driven pure tensile load as shown in Figure 1. The Neo-Hookean hyperelastic and damage material properties are reported in Table. 2. Assume that the healing process is beginning from time $t=100 \ days$, and $Q\left( {t'} \right) = {e^{- {\eta _t}t'}}$  with $t'=t-{\rm{100}}$, where parameter ${\eta _t}$ describes the speed of mass degradation. {According to \cite{Comellas2016}, a relatively large bulk modulus is chosen compared with the shear modulus.}
	
	\subsubsection{G$\&$R constants model}
	\label{5.1.1}
	The performance of \emph{G}\&\emph{R constant model} is firstly tested by calculating the variation of average Cauchy stress along the right side of plate with time as shown in Figure 2. Figure 2(a) shows the influence of the degradation speed parameter ${\eta _t}$ in the \emph{G}\&\emph{R constant model}, where a largrer ${\eta _t}$ causes a higher stress, meaning that a faster degradation (mass decrease) leads to a higher level of healing. Figure 2(b) investigates the influence of the healing fraction $k_g$. The following values are tested: $k_g=0.8$, $k_g=1.0$ and $k_g=1.5$, respectively. Firstly, the \emph{G}\&\emph{R constant model} simulates the increase of stress in the process of healing. {Secondly, low $k_g$ values cause higher stress, as both $k_g<1$ and $\dot Q(t) < 0$ will lead $\dot \rho (s) < 0$ according to Equation (13), so a smaller inelastic deformation ${\bf F} _g$ is produced in Equation (9), and as a constant total displacement loading is applied, a larger elastic deformation ${\bf F} _e$ is obtained in Equation (3), finally the obtained stress is higher than the one with $k_g\ge 1$.}
	
	\subsubsection{G$\&$R homeostatic model}
	\label{sec.5.1.2}
	Figure 2(c) and Figure 2(d) illustrates the stress curves for the \emph{G}\&\emph{R homeostatic model}, where the influence of the homeostatic stress value $\bm{\sigma}_h$  and of the gain parameter $k_{\sigma}$ is shown in Figure 2(c) and Figure 2(d), respectively. {Considering that the homeostatic stress was consistent with stresses commonly applied to soft tissues in vivo \cite{Humphrey2014,Cyron2014}, here we set the values of homeostatic stresses lower that the maximum stress reached after damage, so that G$\&$R worked at reducing this stress in order to converge towards the homeostatic stress, consequently inducing healing.} Results show the convergence of the stress towards the homeostatic stress after healing for all the tested cases. A larger gain parameter $k_\sigma$  results in a faster convergence. {Furthermore, the sensitivity of the size of time step $\Delta t $ is reported in Figure 2(e), the results show that different $\Delta t $ values have no significant influence on the convergence towards the homeostatic state.}
	
	\subsubsection{Comparison of the G$\&$R models with the non-local Comellas model}
	\label{sec.5.1.3}
	The $G\&R \ constant \ model$, the $G\&R \ homeostatic \ model$ and the \emph{non-local Comellas model} are compared in the Figure 3. {It is shown that both the $G\&R \ constant \ model$ and the $G\&R \ homeostatic \ model$ yield a non-zero stress as the displacement loading is entirely unloaded as shown in Figure 3(a). Accordingly, the temporal variations of elastic and inelastic deformations are shown in Figure 3(b), in which nonzero inelastic deformations obtained in the $G\&R \ constant \ model$ and the $G\&R \ homeostatic \ model$ are shown. Comparatively, there is no inelastic deformation for the $non$-$local \ Comellas \ model$.}
	
	\subsection{Open-hole plate }
	\label{sec.5.2}
	The second numerical example is a rectangular plate with a hole, loaded under displacement-driven conditions. The geometry and the loading curves are shown in Figure 4. The Neo-Hookean hyperelastic and damage material properties are reported in Table 3. Due to the symmetry, only a quarter of the plate is analyzed. {For the material parameters, as in Example 1, a relatively large bulk modulus is chosen compared with the shear modulus \cite{Comellas2016}.}
	
	\subsubsection{G$\&$R constant model}
	\label{sec.5.2.1}
	The stress curves shown in Figure 12(a) prove the mesh-independence for the \emph{G}\&\emph{R constant model}. The evolution of the time-dependent damage function  $H(t)$ is shown thoughout the healing process in Figure 5 for two different mesh sizes. Again, the results are fully mesh-independent.
	
	{The influence of non-local effects of the $G\&R \ constant \ model$ is investigated in Figure 6 with different $c_d$ values. $c_d$ is the parameter representing the effect of internal length scales, or more specifically as the parameter related to the average length of microstructural components in the soft tissues. The results show that a higher $c_d$ leads to smaller damage and a narrower damage zone during both the damage and healing process. This can be explained by the larger width of the ‘activated zone’ in case of higher $c_d$ in the model \cite{Dimitrijevic2008,Dimitrijevic2011}, hence changes of the damage variable will affect a larger region, which is related to larger internal length scales of the soft tissue. }
	
	\subsubsection{G$\&$R homeostatic model}
	\label{sec.5.2.2}
	For the \emph{G}\&\emph{R homeostatic model}, the stress curves and the evolution of damage distribution with time are shown in Figure 12(b) and Figure 7 respectively, again revealing a very good mesh-independence, the homeostatic stress is set to a value of $0.00815\ $MPa for the average Cauchy stress at the right side of plate, which represents stress at time $t=40 \ days$. {Unlike the previous results obtained with the $G\&R \ constant \ model$, the stress decreases and tends to converge toward a target value during healing, which represents the homeostatic state. }
	
	{The influence of non-local effects in the $G\&R \ homeostatic \ model$ is also investigated in Figure 8. The effect of internal length scales is shown with different $c_d$ values. The decrease of the damage region is shown during healing with a larger value of $c_d$.} 
	
	Figure 9 illustrates the influence of values of homeostatic stress and the level of damage, respectively. In Figure 9(a), the results show good convergence to the state of homeostatic stree for healing with three different prescribed $\bm{\sigma}_h$. Moreover, the ability of the \emph{G}\&\emph{R homeostatic model} to simulate different levels of damage/healing is analyzed by varying the penalty parameter $\beta_d$  in the non-local damage model as shown in Figure 9(b). It is shown that although the \emph{G}\&\emph{R homeostatic model} is capable of simulating the process of convergence of stress to the homeostatic state for $\beta_d=0.004$ and $\beta_d=0.006$, a simulation failure occurs for $\beta_d=0.002$ with more damage contained. The reason of computational failure could be that a more severe damage causes a more inhomogeneous concentrated stress field, this could cause some numerical difficulty in the computation of healing by coupling the numerical approximation in both spatial and time domains.
	
	\subsubsection{Non-local Comellas model}
	\label{sec.5.2.3}
	Similarly, the response obtained with the $non$-$local \ Comellas \ model$ is shown in Figure 10, Figure 11 and Figure 12(c). The average stress curves and the distribution of damage fields $H(t)$ show that there is no mesh dependence, the non-local approach has permitted to overcome the mesh-dependence which was reported in Comellas et al. \cite{Comellas2016}. {Similar results as for the $G\&R \ healing \ models $ are found for the $Comellas \ model$ by using different $c_d$ values, as shown in Figure 11.}
	\subsection{Balloon angioplasty case study}
	\label{sec.5.3}
	The third case study is related to damage induced by balloon angioplasty and its healing for a long-time scale. The two-dimensional geometry shown in Figure 13(a) was previously established by Badel et al. \cite{Badel2014}, inspired from histological pictures of epicardial coronary arteries from Viles-Gonzalez et al. \cite{Viles-Gonzalez2011}. The coronary artery is assumed to consist of a single medial layer containing an atherosclerotic plaque, and the balloon used for angioplasty is modeled as a thin circular structure whose diameter increases during the angioplasty process. The medial layer and the plaque are modeled based on a Neo-Hookean hyperelastic model, and the balloon is modeled with a linear elastic model. The geometry and the FEM mesh are shown in Figure 13(b), and the material parameters are reported in Table 4.
	
	The only boundary conditions to be assigned are the nodal displacements of the balloon. A radial displacement is imposed to each node from its initial position, ${d_i}=0.5\ mm$, to give a final deformed diameter, ${d_f}=1.0\ mm$. In the following, we use variable $\lambda={(d_c-d_i)}/{(d_f-d_i)}$, where  $d_c$ is the current diameter of the balloon, as a gauge of the inflation progress. The balloon inflation is applied from time $t=0$ to $t=100 \ days$, and the healing is set to begin from time $t=100 \ days$ and the boundary condition is set as constant. {Note that all the degrees of freedom of the balloon are prescribed as Dirichlet boundary conditions. Therefore, the response of the balloon is completely independent of the material behavior assigned to it, so we assigned a linear elastic model for the balloon.}
	
	The process from damage to healing is simulated during balloon angioplasty for the same three models as in previous sections. The final deformation of balloon is set as ${d_f}=1.0\ mm$  for all three models. Results are shown in Figure 14. All three models successfully simulated medial healing after damage, although the results are slightly different in the distribution of damage fields. Comparatively, the effects of healing are more pronounced for the \emph{G}\&\emph{R constant model} and for the \emph{non-local Comellas model}, but the \emph{G}$\&$\emph{R homeostatic model} shows a more stable process due to the homeostatic condition.
	
	Figure 15 illustrates the effect of different level of balloon inflation setting three different diameters (a)${d_f}=0.9 \ mm$, (b)${d_f}=1.0 \ mm$ and (c)${d_f}=1.1 \ mm$ by using the \emph{G}\&\emph{R constant model}. The first and second column shows two damage fields $H(t)$ in the damage process during the balloon dilation, illustrating that more damage is induced under a larger dilation size. The third and fourth column show again the two damage fields $H(t)$ throughout the healing process, in which a recoverable damage can be observed, and an obvious change of geometry of the media layer can be found for case (c) at time  $t=200 \ days$. This shows the ability of the proposed model in simulating the healing process along with the induced growth deformation.
	
	\section{Conclusions}
	\label{sec:6}
	We have developed two new gradient-enhanced continuum healing models for soft tissues, including the gradient-enhanced G$\&$R healing model and the gradient-enhanced version of the healing model proposed by Comellas et al. \cite{Comellas2016} using Abaqus with UEL, and we have shown their potential for applied problems.
	
	A first advantage of the two healing models is their {ability} to simulate the healing process non-locally by introducing the gradient-enhanced variable. Numerically, a good mesh independence is achieved in the simulation of healing, even when damage is concentrated in a narrow region. 
	
	For the gradient-enhanced $G\&R \ healing \ model$, the time-dependent inelastic growth is introduced into the conventional gradient-enhanced damage model to describle the process of G$\&$R in healing in the framework of the temporally homogenized growth model \cite{Cyron2016}. In this paper, two approaches to determine the rate of mass production are discussed, including the $G\&R \ constant$ $ model$ and the $G\&R \ homeostatic \ model$, and the growth direction are determined according to local principal stress directions. As shown in numerical examples where the effects of the G$\&$R parameters on results are discussed, it seems that the level of damage before the beginning of healing could be a sensitive factor for the convergence towards homeostasis.
	
	Moreover, the difficulty of mesh dependence in original Comellas healing model \cite{Comellas2016} has been well overcome by virtue of the gradient-enhanced term. Nevertheless, the gradient-enhanced Comellas have not considered the influence of inelastic growth deformation in healing, in comparison with the gradient-enhanced $G\&R \ healing \ models.$
	
	Aiming to approach the applied problems, healing after damage in balloon angioplasty is simulated by the proposed models in the last numerical example, and the influence of the inflation diameter on healing is investigated. The proposed models have shown good potential for approaching the healing for damaged soft tissues. 
	
	
	The present model is limited to 2D cases and to isotropic hyperelastic models. However, as collagen fibers are essential in healing of soft tissue, the development of a 3D anisotropic model is currently under progress in order to address more realistic applications. Besides, the use of UEL presents some limitations such as the definition of slave surfaces in contact analyses. Therefore, self-contact problems cannot be addressed with the current model.
	
	The determination of material parameters is also an important issue for the applications of the present model. Generally, hyperelastic parameters can be identified from experimental data and an abundant literature exists on this topic \cite{avril2017hyperelasticity}. But the identification of other parameters, relative for instance to internal length scales, such as the gradient parameter $c_d$ and the penalty parameter $\beta_d$ will require inverse analyses to be deduced for practical applications.
	
	In summary, in this manuscript, two gradient-enhanced constitutive healing models for biological soft tissues including non-local variables have been presented. Important developments are currently under progress for considering the anisotropic constitutive and extension to 3D for more practical applications. The development of a 3D anisotropic model will permit simulating arterial healing after surgical procedures such as angioplasty and stent deployment. This will require defining realistic geometries and appropriate constitutive models to be able to predict the long-term adaptation of arteries to these invasive procedures. Available information about the microstructure of concerned arteries will permit defining the internal length scales.
	
	\begin{acknowledgements}
		The research leading to this paper is funded by ERC-2014-CoG-BIOLOCHANICS [647067], NSFC-ERC grant [11711530644], NSFC grant [11572077], Open Fund from the State Key Laboratory of Structural Analysis for Industrial Equipment [GZ1708]. The authors also acknowledge the support from Prof. Pierre Badel for the model of balloon angioplasty
	\end{acknowledgements}
	
	\clearpage
	\bibliographystyle{spphys}
	\bibliography{MyCollection}

	\clearpage
	\listoftables
	
	\clearpage
	\begin{table}
		\caption{Different steps of the numerical implementation at the Gauss point level for gradient-enhanced continuum healing models.}
		\label{tab:1}       
		\begin{tabular}{l}
			\noalign{\smallskip}\hline	
			0. 	$Initialization \ at \ t=0 \ and \ n=0 $  \\
			\quad Mechanical damage $d^n=0$  and healing function $H^n=0$ \\
			1.	$Algorithm \ at \ each \ load \ increment \ n$ \\
			\quad 1.1	Given: Total deformation gradient tensor ${\bf F}$, elastic deformation ${\bf F}_e$ , inelastic deformation ${\bf F}_g$ and material properties
			\\
			\quad 1.2	Compute driving force from Equation (29)\\
			\quad 1.3	Check damage condition from Equation (30), if ${\rm {\Phi _d}}{\rm{= }}q - {r_1} \le 0$ go to 1.5, else go to 1.4\\
			\quad 1.4	Update damage $d^{n+1}=d^{n}+\Delta d$ from Equation (31)
			\\
			\quad 1.5	Update healing function $H^{n+1}$ \\ 
			\qquad		IF ($G\&R \ constant \ model$)\\
			\qquad		Compute $H^{n+1}$ from Equation (39)\\
			\qquad		IF ($G\&R \ homeostatic \ model$)\\
			\qquad		Compute $H^{n+1}$ from Equation (40)\\
			\qquad		IF (\emph{Non-local Comellas model})\\
			\qquad		Compute $H^{n+1}$ from Equation (42)\\
			\quad 1.6 	Compute the stress state for the present step  $\bm \sigma$
			\\
			\quad 1.7 	Compute tangent moduli $d \bm{\sigma} / d \phi$,$2dy/d \bf{g}$, $dy/d \phi$ and $d {\bf y} /d \nabla_x \phi$ in Equations (34-37).\\
			\noalign{\smallskip}\hline
		\end{tabular}
	\end{table}
	
	\clearpage
	\begin{table}
		\caption{Hyperelastic, damage and healing material parameters used in the homogeneous uniaxial tensile test example.}
		\label{tab:2}       
		\begin{tabular}{lllll}
			\hline\noalign{\smallskip}
			Type & Description & Symbol  & Values &Units\\
			\noalign{\smallskip}\hline\noalign{\smallskip}
			\multirow{2}{*}{Hyperelastic}	 & Shear modulus	 & $\mu_e$	 	& 1.5	 & MPa\\
			& Bulk modulus		 & $\kappa_e$ 	& 75.0	 & MPa\\
			\noalign{\smallskip}\hline
			\multirow{2}{*}{Damage}      	 & Saturation parameter & $\eta_d$ & 1.0 & MPa$^{-1}$\\
			& Damage threshold          	 & $\kappa_d$ & 8.0 & MPa\\
			\noalign{\smallskip}\hline
			\multirow{6}{*}{Healing}      & \multirow{2}{*}{\emph{G}\&\emph{R constant model} } & $k_g$ &$[0.8,1.0,1.5]$& $-$\\
			&  & $\eta_t$ &$[0.001,0.005,0.01]$& $-$\\
			& \multirow{2}{*}{$G\&R \ homeostaic \ model$} & $k_{\sigma}$& $[0.005,0.001,0.02]$& $-$\\
			&  & $\eta_t$ &0.01& $-$\\
			& \multirow{2}{*}{\emph{non-local Comellas model}} & $\dot{\eta}$ & 0.01& $days^{-1}$\\
			&  & $\xi$ & 0.0 & $-$\\
			\noalign{\smallskip}\hline
		\end{tabular}
	\end{table}
	
	\clearpage
	\begin{table}
		\caption{Hyperelastic, damage and healing material parameters used in the open-hole tensile test example.}
		\label{tab:3}       
		\begin{tabular}{lllll}
			\hline\noalign{\smallskip}
			Type & Description & Symbol  & Values &Units\\
			\noalign{\smallskip}\hline\noalign{\smallskip}
			\multirow{2}{*}{Hyperelastic}	 & Shear modulus	 & $\mu_e$	 	& 0.1	 & MPa\\
			& Bulk modulus		 & $\kappa_e$ 	& 5.0	 & MPa\\
			\noalign{\smallskip}\hline
			\multirow{2}{*}{Damage}      	 & Saturation parameter & $\eta_d$ & 1.0 & MPa$^{-1}$\\
			& Damage threshold          	 & $\kappa_d$ & 0.002 & MPa\\
			\noalign{\smallskip}\hline
			\multirow{3}{*}{Regularisation} &Degree of regularisation& $c_d$ &{[0.1,1.0,10]}& MPa$\cdot mm^2$\\
			& Penalty parameter & $\beta_d$ & $[0.002,0.004,0.006]$ & MPa\\
			& (Non-)local switch & $\gamma_d$ & $1.0$ & $-$\\
			\noalign{\smallskip}\hline
			\multirow{6}{*}{Healing}      & \multirow{2}{*}{\emph{G}\&\emph{R constant model}} & $k_g$ &$[0.8,1.0,1.5]$& $-$\\
			&  & $\eta_t$ &$[0.001,0.005,0.01]$& $-$\\
			& \multirow{2}{*}{$G\&R \ homeostaic \ model$} & $k_{\sigma}$& $[0.005,0.001,0.02]$& $-$\\
			&  & $\eta_t$ &0.01& $-$\\
			& \multirow{2}{*}{\emph{non-local Comellas model}} & $\dot{\eta}$ & 0.01& $days^{-1}$\\
			&  & $\xi$ & 0.0 & $-$\\
			\noalign{\smallskip}\hline
		\end{tabular}
	\end{table}
	
	\clearpage
	\begin{table}
		\caption{Material parameters used in the balloon angioplasty case study \cite{Badel2014}.}
		\label{tab:4}       
		\centering
		\begin{tabular}{llllll}
			\hline\noalign{\smallskip}
			Type & & Description  & Symbol  & Values &Units\\
			\noalign{\smallskip}\hline\noalign{\smallskip}
			\multirow{6}{*}{Hyperelastic}	 & \multirow{2}{*}{Medial layer} & Shear modulus & $\mu_e$ & 200	 & kPa\\
			& & Bulk modulus		 & $\kappa_e$ 	& 2.0	 & MPa\\
			
			& \multirow{2}{*}{Plaque} & Shear modulus & $\mu_p$ & 20	 & kPa\\
			& & Bulk modulus		 & $\kappa_p$ 	& 34	 & kPa\\
			
			& \multirow{2}{*}{Balloon} & Shear modulus & $\mu_b$ & 0.5	 & MPa\\
			& & Bulk modulus		 & $\kappa_b$ 	& 2.0	 & MPa\\
			\noalign{\smallskip}\hline
			
			\multirow{2}{*}{Damage}   &  \multirow{2}{*}{Medial layer}   & Saturation parameter & $\eta_d$ & 1.0 & MPa$^{-1}$\\
			& & Damage threshold          	 & $\kappa_d$ & 5.0 & kPa\\
			\noalign{\smallskip}\hline
			
			\multirow{3}{*}{Regularisation} & \multirow{3}{*}{Medial layer} & Degree of regularisation& $c_d$ &1.0& MPa$\cdot mm^2$\\
			&& Penalty parameter & $\beta_d$ & 5.0 & kPa\\
			&& (Non-)local switch & $\gamma_d$ & $1.0$ & $-$\\
			\noalign{\smallskip}\hline
			
			\multirow{6}{*}{Healing}   & \multirow{2}{*}{\emph{G}\&\emph{R constant model}} & Gain parameter & $k_g$ & 1.0 & $-$\\
			&  & Survive function parameter & $\eta_t$ & -0.001& $-$\\
			& {$G\&R \ homeostaic \ model$} & Gain parameter & $k_{\sigma}$& 0.05 & $-$\\
			& \multirow{2}{*}{\emph{non-local Comellas model}}& Healing rate & $\dot{\eta}$ & 0.015& $days^{-1}$\\
			&  &Un-recover percentage & $\xi$ & 0.0 & $-$\\
			\noalign{\smallskip}\hline
		\end{tabular}
	\end{table}

	\clearpage
	\listoffigures
	\clearpage
	\begin{figure}
		\centering
		\includegraphics[scale=0.8]{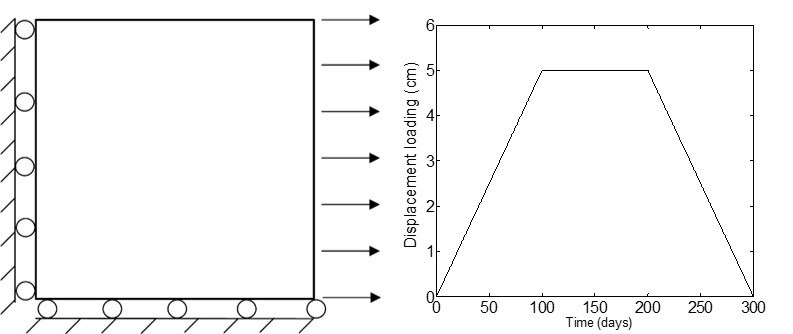}
		\caption{Geometry and displacement applied for the uniaxial tension case study}
		\label{fig:1}       
	\end{figure}
	
	\clearpage
	\begin{figure}
		\centering
		\includegraphics[scale=0.5]{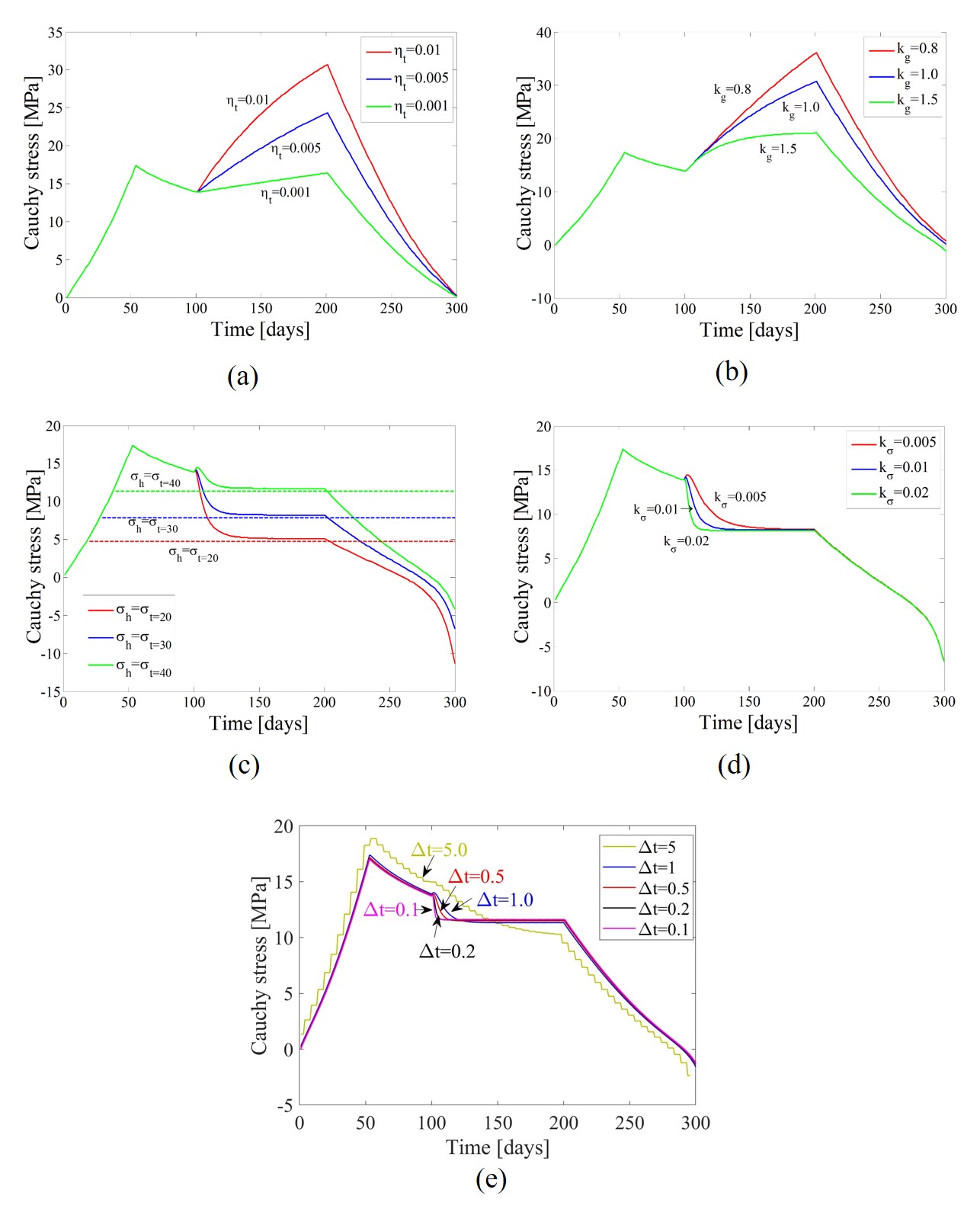}
		\caption{Stress curves obtained in the homogeneous uniaxial tension example. (a) Influence of the degradation speed parameter $\eta_t$ for $G\&R \ constant \ model$. (b) Influence of the healing parameter $k_g$ for $G\&R \ constant \ model$. (c) Influence of the homeostatic stress $\bm \sigma_h$ for $G\&R \ homeostatic \ model$. (d) Influence of the gain parameter $k_{\sigma}$ for $G\&R \ homeostatic \ model$. {(e) Influence of time step size for the $G\&R \ homeostatic \ model.$}}
		\label{fig:2}       
	\end{figure}
	
	\clearpage
	\begin{figure}
		\centering
		\includegraphics[scale=0.5]{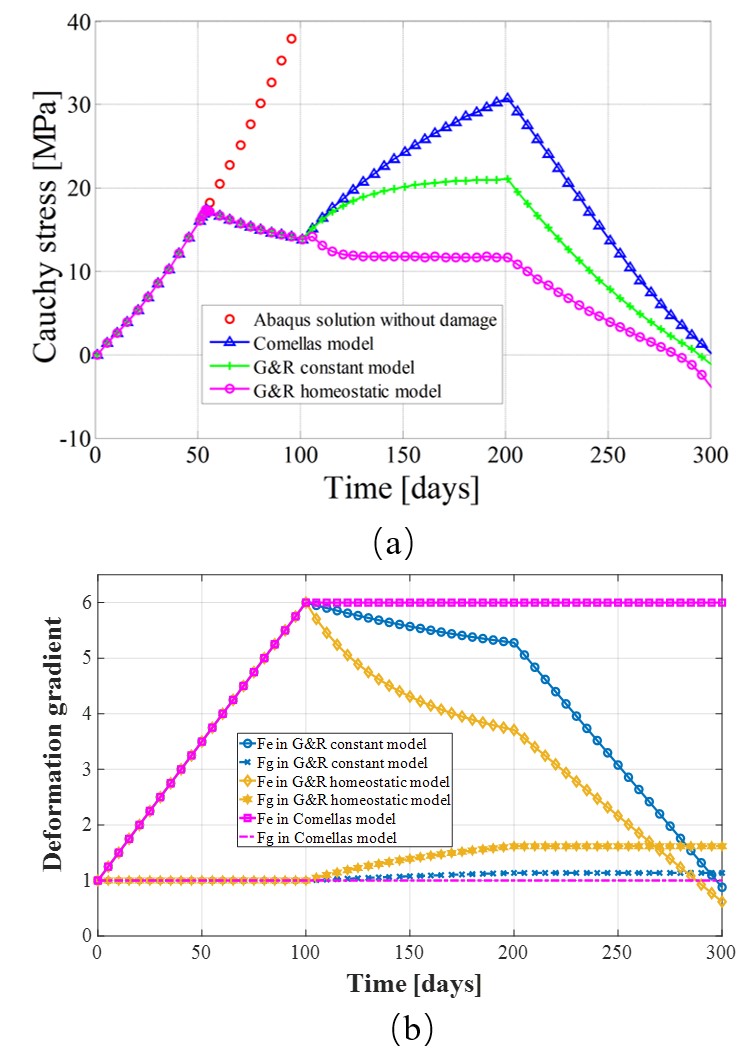}
		\caption{Comparison of $G\&R \ constant \ model$, the $G\&R$ \emph{homeostatic model} and the $non$-$local \ Comellas \ model.$ (a) Stress-time curves. {(b) Temporal variations of the deformation gradient.}}
		\label{fig:3}       
	\end{figure}
	
	\clearpage
	\begin{figure}
		\centering
		\includegraphics[scale=0.8]{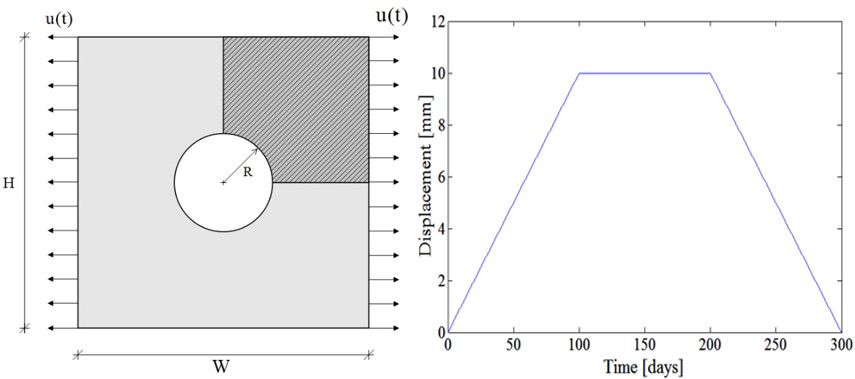}
		\caption{Geometry and displacement applied for the open-hole rectangular plate case study.}
		\label{fig:4}       
	\end{figure}
	
	\clearpage
	\begin{figure}
		\centering
		\includegraphics[scale=0.8]{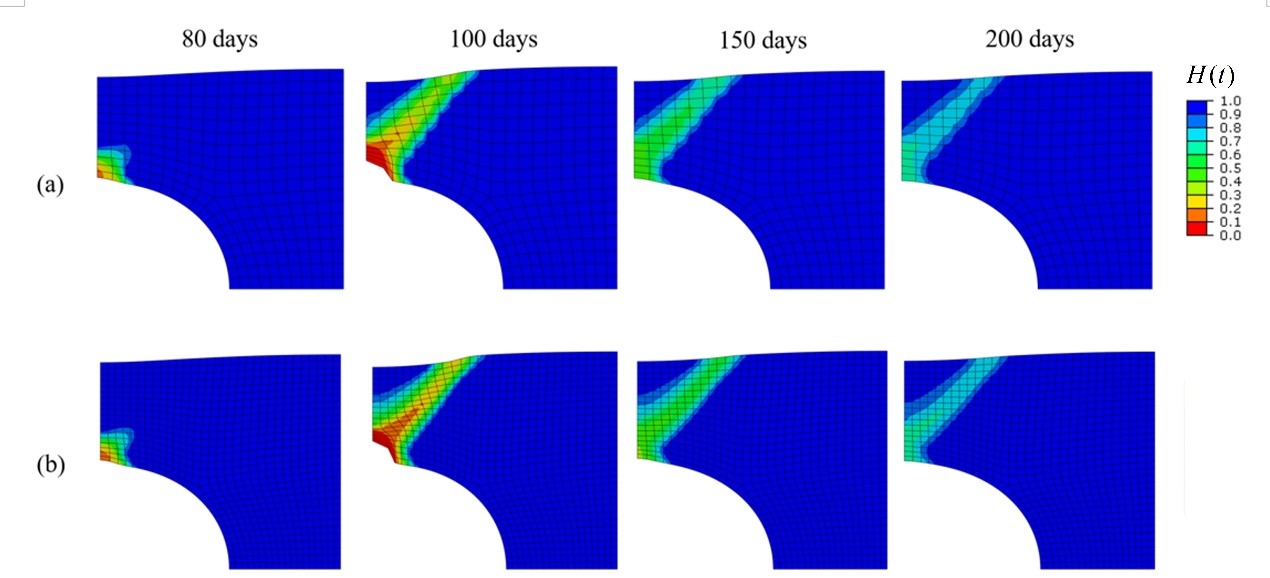}
		\caption{Evolution of the damage fields throughout the healing process for the $G\&R \ constant \ model.$ (a) Results with a coarse mesh of 286 elements. (b) Results obtained with a fine mesh of 793 elements.}
		\label{fig:5}       
	\end{figure}
	
	\clearpage
	\begin{figure}
		\centering
		\includegraphics[scale=0.35]{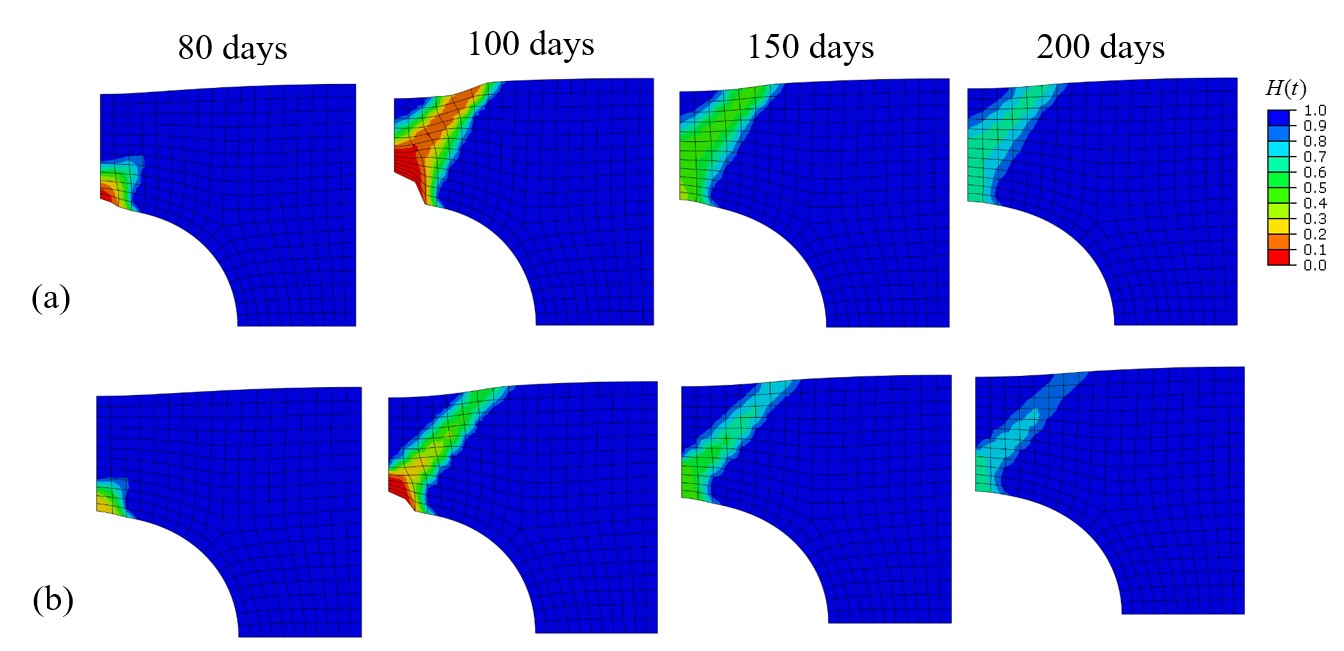}
		\caption{Evolution of the damage fields throughout the healing process for the $G\&R \ constant \ model.$ (a) Results with $c_d=0.1$. (b) Results with  $c_d=10$.}
		\label{fig:6}       
	\end{figure}
	
	\clearpage
	\begin{figure}
		\centering
		\includegraphics[scale=0.4]{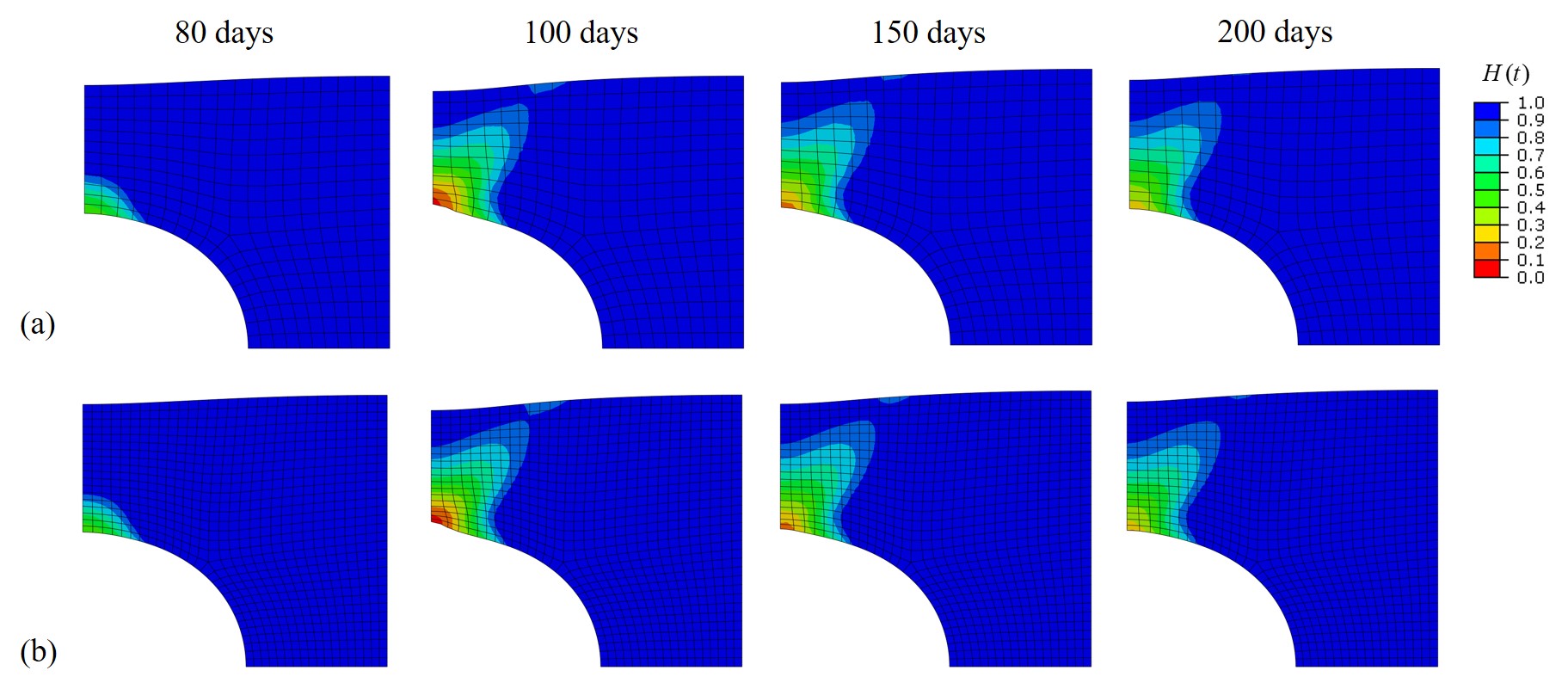}
		\caption{Evolution of the damage fields throughout the healing process for the $G\&R \ homeostatic  \ model.$ (a) Results with a coarse mesh of 286 elements. (b) Results obtained with a fine mesh of 793 elements.}
		\label{fig:7}       
	\end{figure}
	
	\clearpage
	\begin{figure}
		\centering
		\includegraphics[scale=0.3]{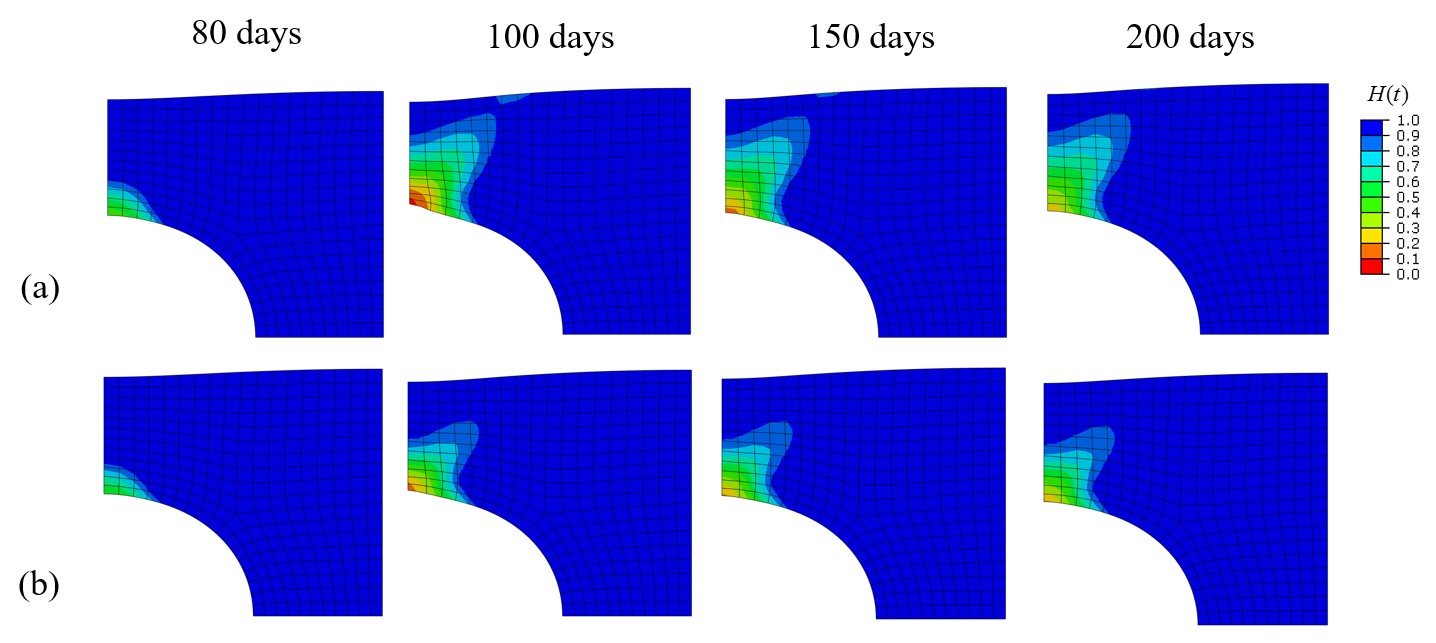}
		\caption{Evolution of the damage fields throughout the healing process for the $G\&R \ homeostatic \ model.$ (a) Results with $c_d=1.0$. (b) Results with $c_d=10$.}
		\label{fig:8}       
	\end{figure}
	
	\clearpage
	\begin{figure}
		\centering
		\includegraphics[scale=0.8]{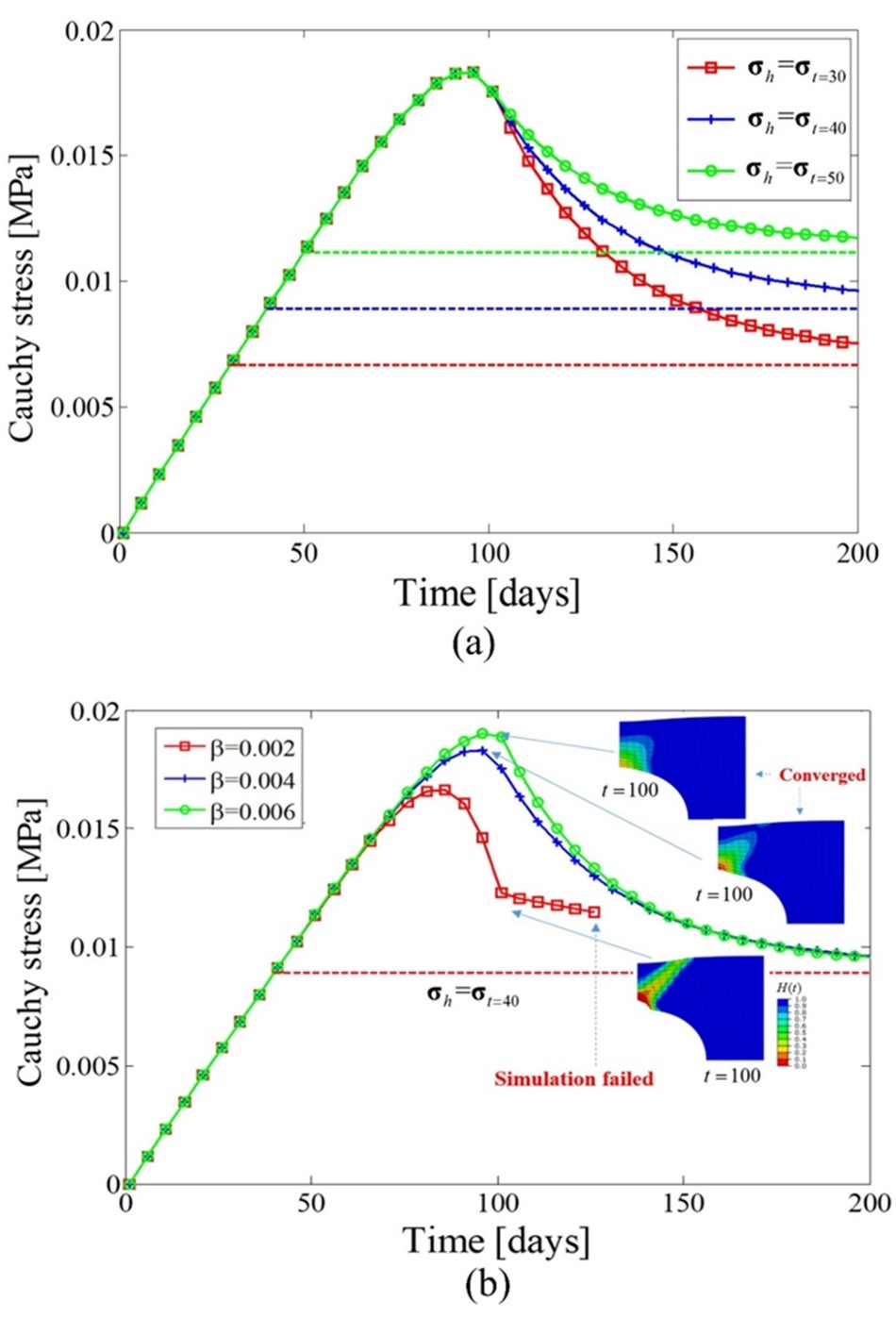}
		\caption{Evolution of the damage fields throughout the healing process for the $G\&R \ homeostatic \ model$ (a) Results with different homeostatic stress values. (b) Results with different levels of damage/healing.}
		\label{fig:9}       
	\end{figure}
	
	\clearpage
	\begin{figure}
		\centering
		\includegraphics[scale=0.4]{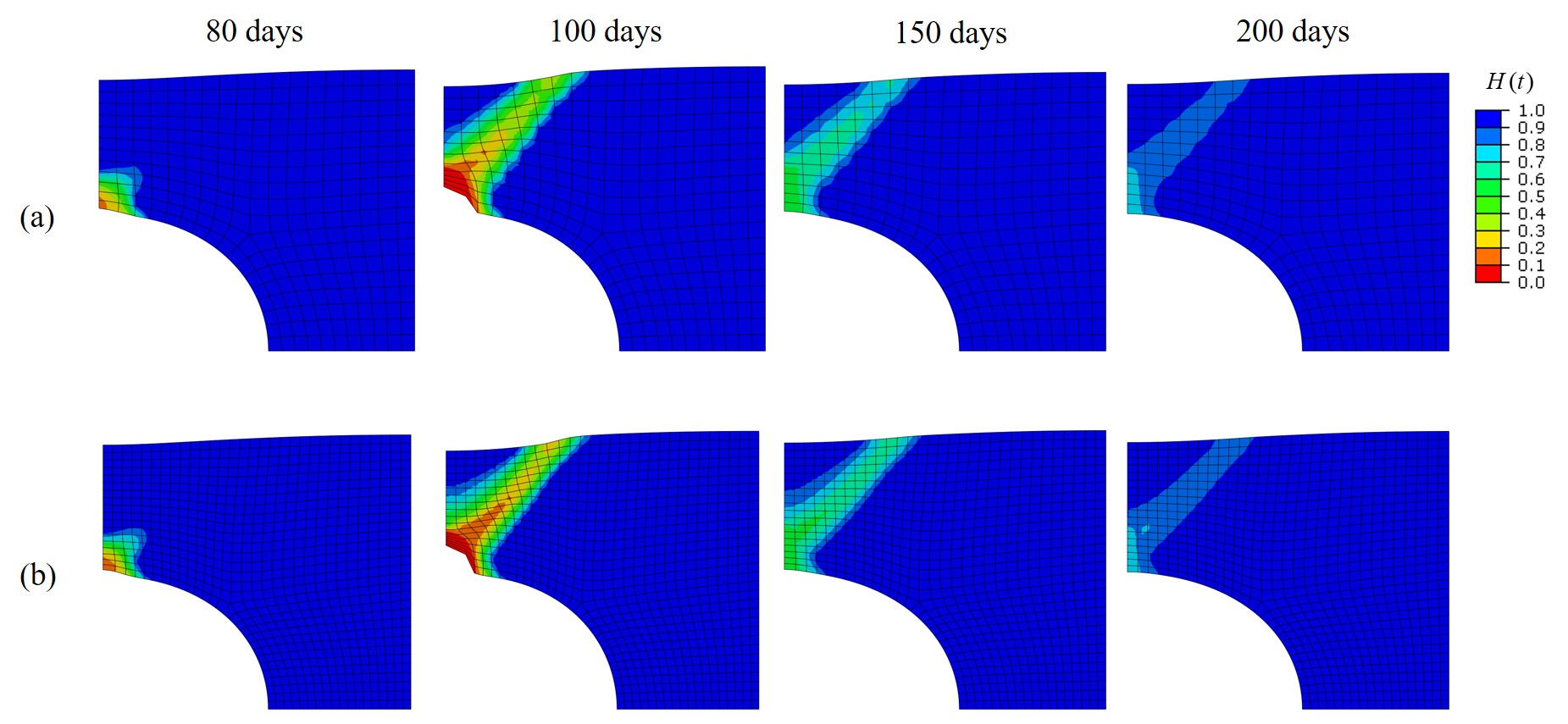}
		\caption{Evolution of the damage fields throughout the healing process for the $non$-$local \ Comellas \ model.$ (a) Results with a coarse mesh of 286 elements. (b) Results obtained with a fine mesh of 793 elements.}
		\label{fig:10}       
	\end{figure}
	
	\clearpage
	\begin{figure}
		\centering
		\includegraphics[scale=0.35]{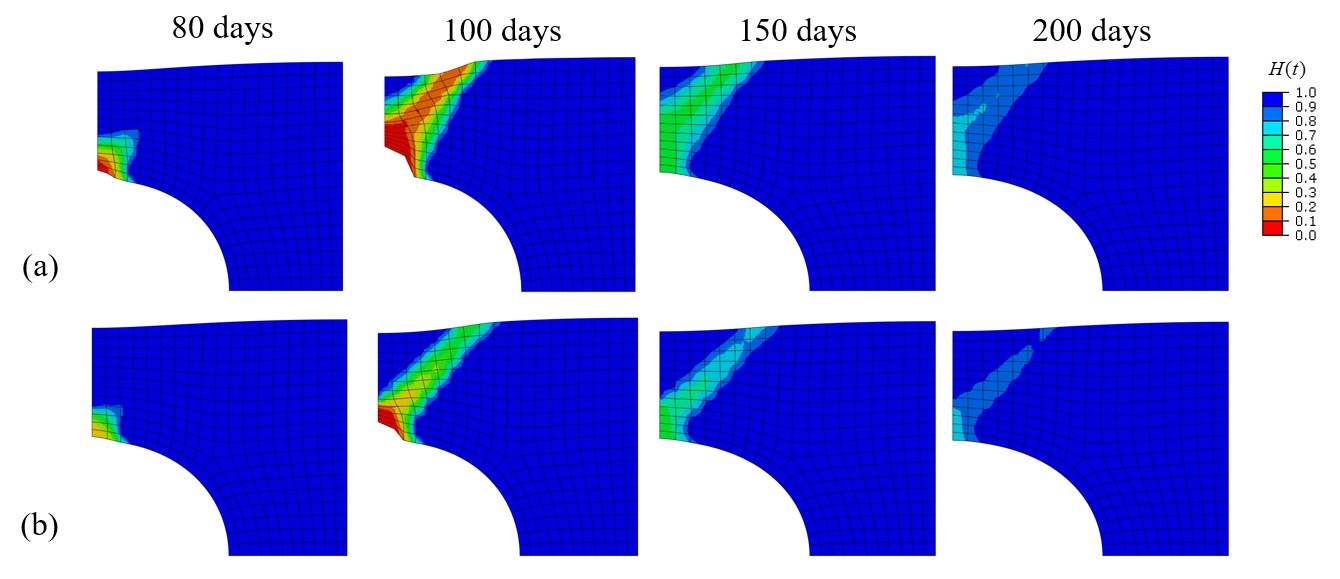}
		\caption{Evolution of the damage fields throughout the healing process for the $non$-$local \  Comellas  \ model.$ (a) Results with $c_d=1.0$. (b) Results with $c_d=10$.}
		\label{fig:11}       
	\end{figure}
	
	\clearpage
	\begin{figure}
		\centering
		\includegraphics[scale=0.45]{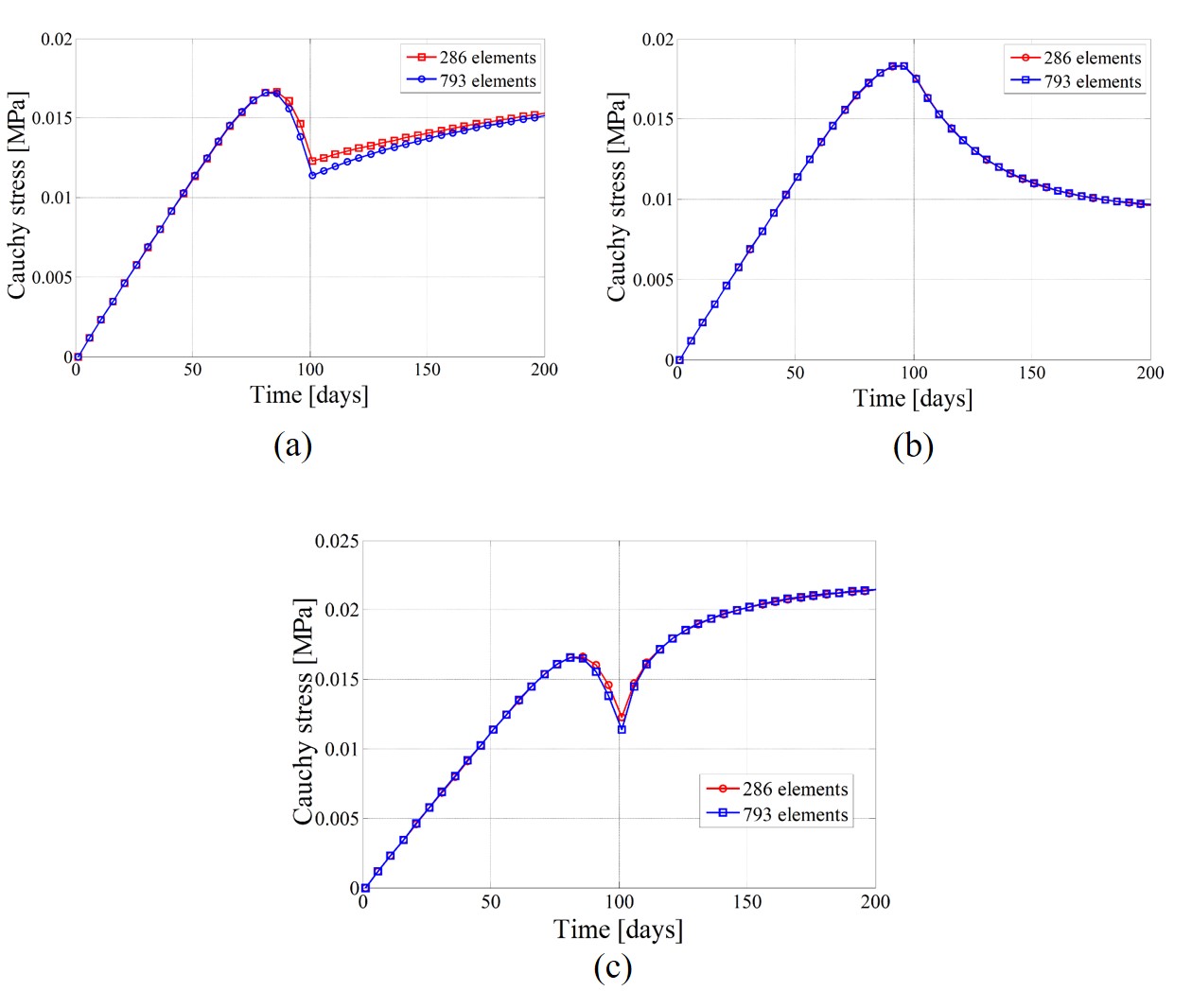}
		\caption{Plate with a hole. Average cauchy stress curves for 286 and 793 elements of three different non-local models. (a) $G\&R$ \emph{constant model}. (b) $G\&R$ $homeostatic \ model$. (c) $Comellas \ model.$}
		\label{fig:12}       
	\end{figure}
	
	\clearpage
	\begin{figure}
		\centering
		\includegraphics[scale=0.6]{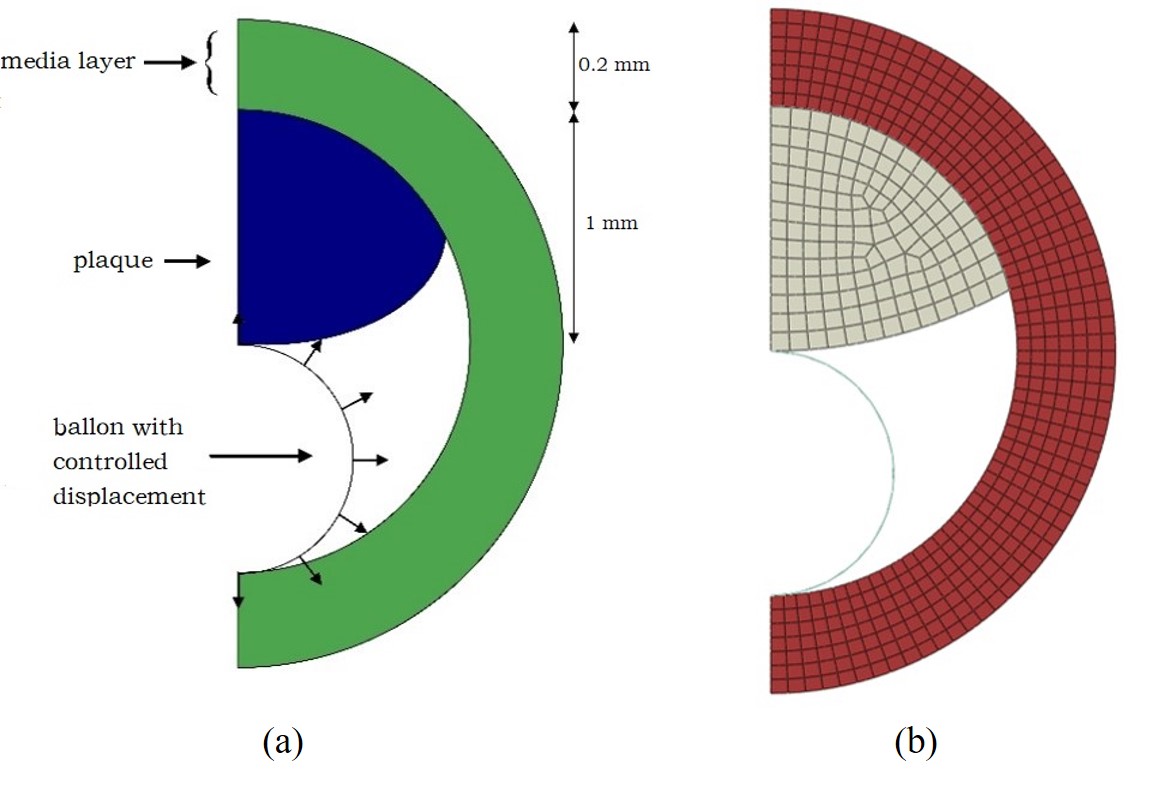}
		\caption{(a) Reference model: geometry, dimensions and boundary conditions. (b) FEM mesh.	}
		\label{fig:13}       
	\end{figure}
	
	\clearpage
	\begin{figure}
		\centering
		\includegraphics[scale=0.35]{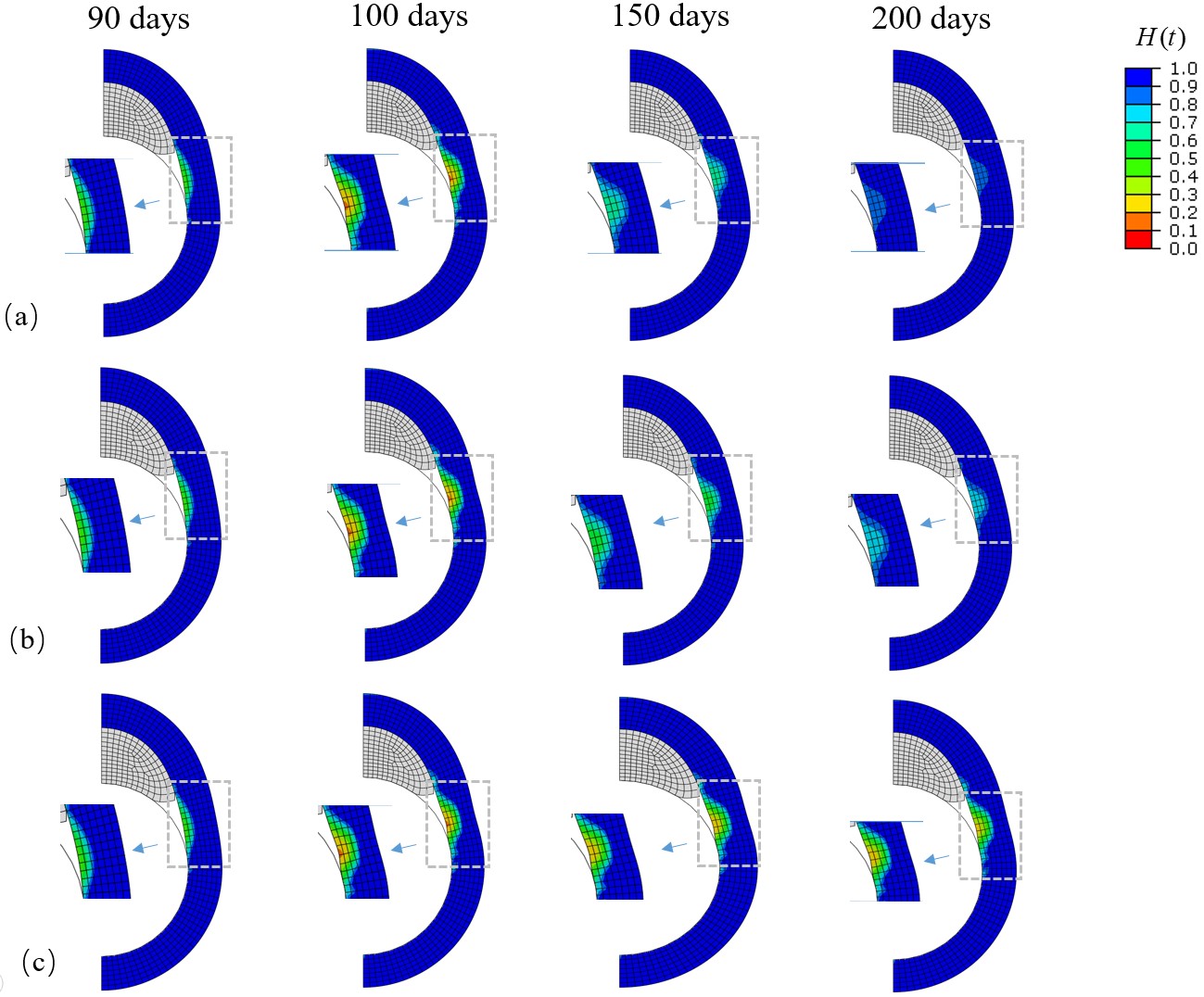}
		\caption{Evolution of the damage fields throughout the healing process for three different non-local models (a) $Comellas\ model$. (b) $G\&R \ constant \ model$. (c) $homeostatic \ model.$}
		\label{fig:14}       
	\end{figure}
	
	\clearpage
	\begin{figure}
		\centering
		\includegraphics[scale=0.35]{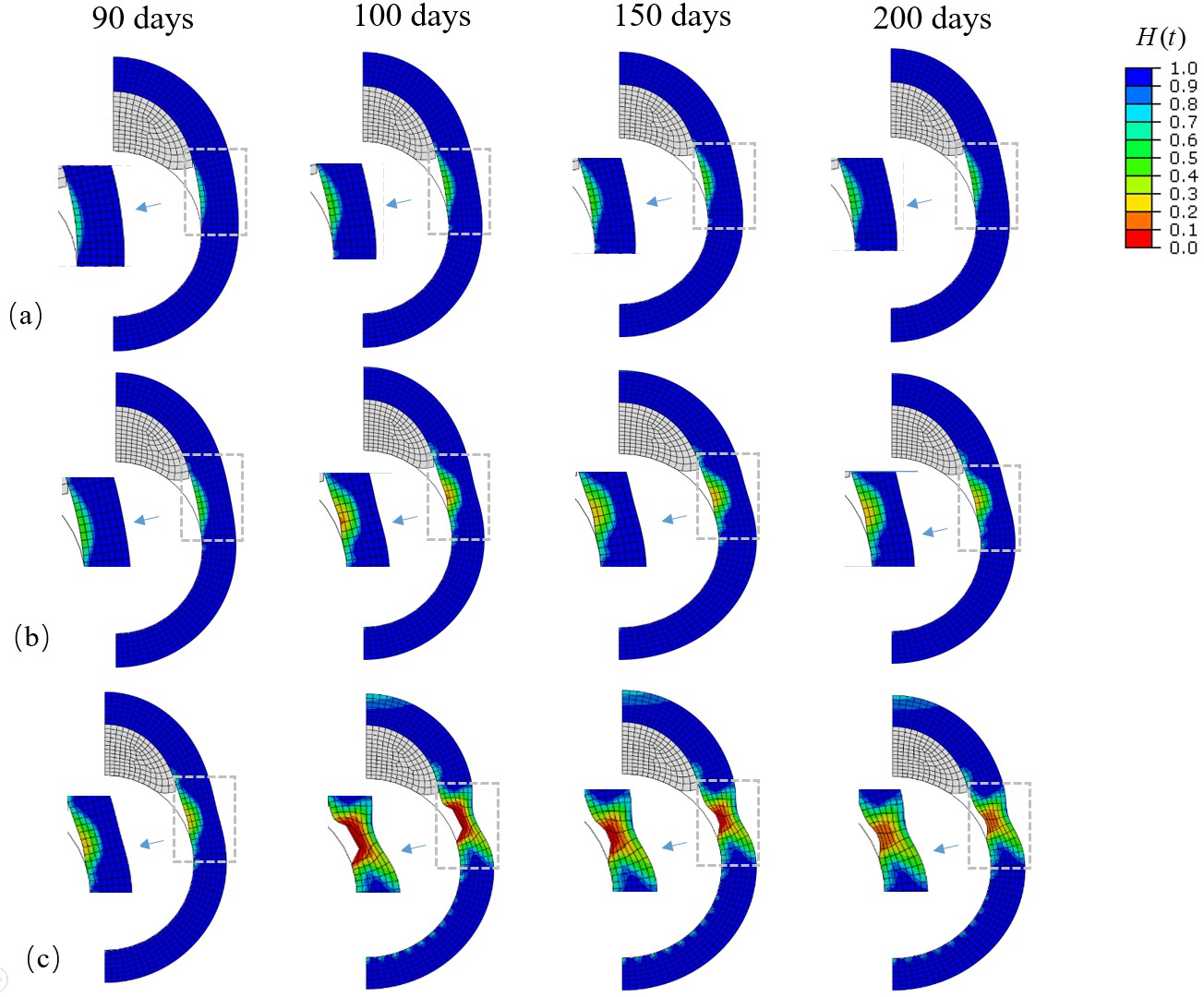}
		\caption{Evolution of the damage fields throughout the healing process for three different inflation diameters (a)$d_f=0.9 \ mm$. (b)$d_f=1.0 \ mm$. (c)$d_f=1.1 \ mm$.}
		\label{fig:15}       
	\end{figure}
	
\end{document}